\documentclass[sigconf]{acmart}

\AtBeginDocument{%
  \providecommand\BibTeX{{%
    \normalfont B\kern-0.5em{\scshape i\kern-0.25em b}\kern-0.8em\TeX}}}

\usepackage[linesnumbered,ruled,vlined]{algorithm2e}

\usepackage{multirow}
\usepackage{mathtools}
\usepackage{listings}
\usepackage{xcolor}
\usepackage{verbatim}
\usepackage[utf8]{inputenc}
\usepackage{hyperref}
\usepackage{enumitem}
\usepackage[utf8]{inputenc}

\copyrightyear{2025}
\acmYear{2025}
\setcopyright{acmcopyright}
\setcctype{by-nc-nd}
\acmConference[CHI '25]{CHI Conference on Human Factors in Computing Systems}{April 26-May 1, 2025}{Yokohama, Japan}
\acmBooktitle{CHI Conference on Human Factors in Computing Systems (CHI '25), April 26-May 1, 2025, Yokohama, Japan}\acmDOI{10.1145/3706598.3713502}
\acmISBN{979-8-4007-1394-1/25/04}

\begin{document}

\title{GenPara: Enhancing the 3D Design Editing Process by Inferring Users' Regions of Interest with Text-Conditional Shape Parameters}

\author{Jiin Choi}
\orcid{0009-0003-2513-6895}
\affiliation{\institution{Design Informatics Lab, Interior Architecture Design \\ Hanyang University}
\city{Seoul}
\country{Republic of Korea}}
\email{jiin4900@gmail.com}

\author{Seung Won Lee}
\orcid{0000-0001-6435-0833}
\affiliation{\institution{Design Informatics Lab, Interior Architecture Design \\ Hanyang University}
\city{Seoul}
\country{Republic of Korea}}
\email{lswgood0901@gmail.com}

\author{Kyung Hoon Hyun}
\authornote{Corresponding author.}
\orcid{0000-0001-6379-9700}
\affiliation{\institution{Design Informatics Lab, Interior Architecture Design \\ Hanyang University}
\city{Seoul}
\country{Republic of Korea}}
\email{hoonhello@gmail.com}

\renewcommand{\shortauthors}{Choi et al.}
\renewcommand{\shorttitle}{GenPara: Enhancing the 3D Design Editing Process by Inferring Users' Regions of\\ Interest with Text-Conditional Shape Parameters}

\begin{abstract}
In 3D design, specifying design objectives and visualizing complex shapes through text alone proves to be a significant challenge. Although advancements in 3D GenAI have significantly enhanced part assembly and the creation of high-quality 3D designs, many systems still to dynamically generate and edit design elements based on the shape parameters. To bridge this gap, we propose \textit{GenPara}, an interactive 3D design editing system that leverages text-conditional shape parameters of part-aware 3D designs and visualizes design space within the Exploration Map and Design Versioning Tree. Additionally, among the various shape parameters generated by LLM, the system extracts and provides design outcomes within the user's regions of interest based on Bayesian inference. A user study (\textit{N} = 16) revealed that \textit{GenPara} enhanced the comprehension and management of designers with text-conditional shape parameters, streamlining design exploration and concretization. This improvement boosted efficiency and creativity of the 3D design process.

\end{abstract}

\newcommand{\longbar}{
    \tikz[baseline] \draw[thick] (0,0) -- (0,2.5ex);%
}




\begin{teaserfigure}
  \includegraphics[width=\textwidth]{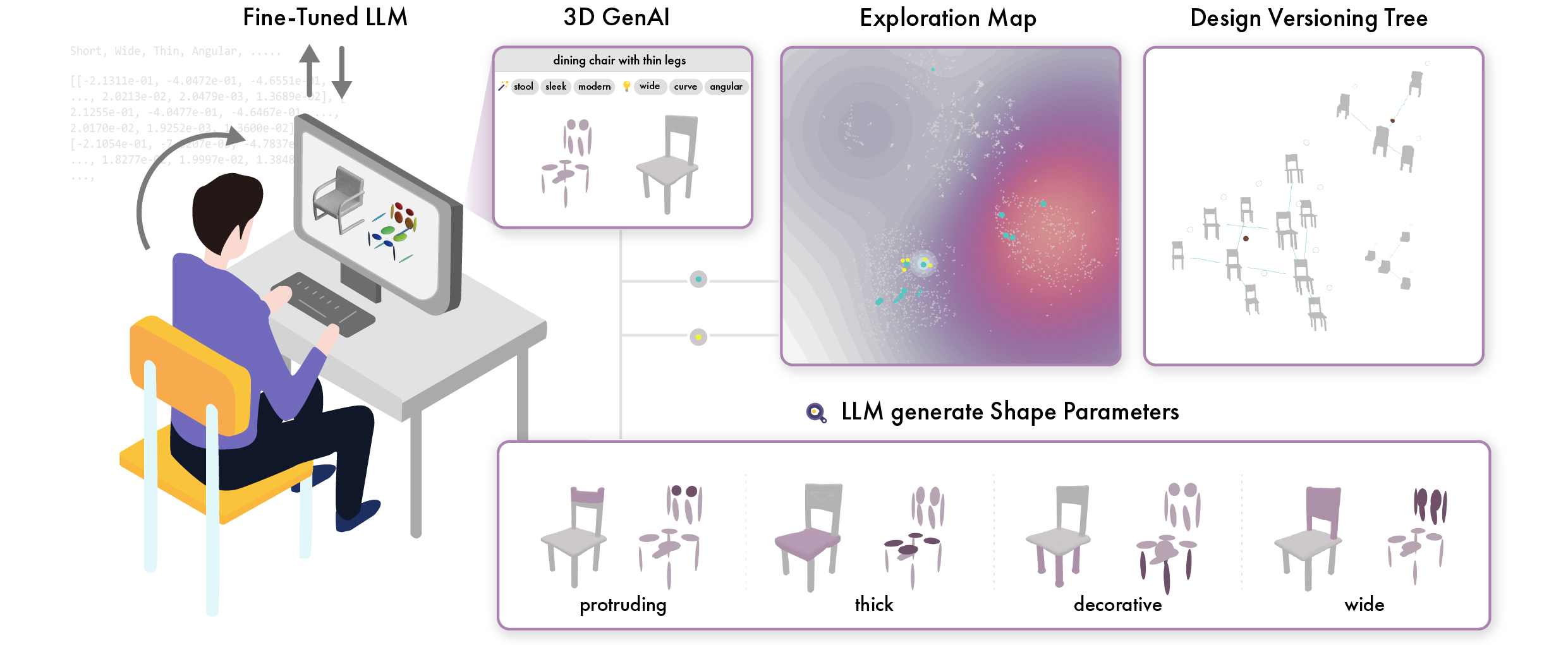}
  \caption{Overview of \textit{GenPara}, an interactive 3D design editing system that leverages text-conditional shape parameters of 3D design and infers the users' regions of interest.}
  \Description{Graphical overview of the \textit{GenPara} system, highlighting its key components. The figure is divided into four sections: 'Fine-Tuned LLM' with a designer at a computer, '3D GenAI' showing labeled chair models, 'Exploration Map' visualizing clustered design parameters, and 'Ontology Map' depicting the relationships among different chair designs using a node-link diagram.}
  \label{fig:teaser}
\end{teaserfigure}
\maketitle

\section{Introduction} 
The design process fundamentally involves designers progressively accumulating their preferences and knowledge, drawing inspiration, and creating new designs \cite{kwon2023understanding,kwon2024comparing,wynn2018process}. Despite the recent advancement of prompt-based generative artificial intelligence (GenAI) systems showing immense promise in various creative fields, effectively utilizing these tools within the 3D design process poses significant challenges. The complexity of design elements often exceeds what can be adequately captured through textual prompts alone \cite{thoring2023augmented,chiarello2024generative}. Designers often find it difficult to articulate the subtle aspects of shapes that are intuitively visual yet complex to express verbally. 

In 3D design, specifying design objectives and visualizing complex shapes through text alone proves to be a significant challenge. This difficulty stems from the the inherent spatial and structural complexities of 3D designs, such as shape transitions of intricate details, layers, parts, and connections \cite{dai2022visual}. These complexities make textual descriptions difficult, as interpretations often vary widely among individuals, hindering standardization and clarity.

State-of-the-art literature dealing with 3D shapes has demonstrated the decomposition and manipulation of 3D shapes at the part level \cite{hertz2022spaghetti,hu2024cns,koo2023salad}, allowing for the synthesis and creation of shape elements based on user input text \cite{sella2023spic} or sketches, optimizing computational and time efficiencies \cite{bandyopadhyay2024sketchinr,bandyopadhyay2023doodle}. The literature focuses on improving shape accuracy, resolution, or part assembly but often overlooks the need for methods that support the nuanced, iterative design exploration process. GenAI-driven design support systems that neglect the nature of design process fail to align with the iterative and nuanced nature of the design process may not reduce their adoption by designers. To address this, GenAI systems should align with the design process, enabling both creative exploration and structured refinement. Accurately comprehending and elucidating specific shape parameters in a way that is  comprehensible by designers is crucial. Moreover, an effective integration of these parameters into the design process supports the concretization of design goals and fosters creative exploration \cite{keurulainen2023amortised,sedlmair2014visual}. These parameters provide a structured way to refine design intent, enabling designers to move beyond static textual prompts and explore the design space iteratively and interactively. Building on these insights, our approach emphasizes the role of text-conditional parameters in linking abstract design goals to tangible outcomes, enabling iterative and structured exploration of complex 3D design spaces. We posit that if 3D GenAI can effectively respond to text-conditional shape parameter of designers, it would enable a robust interactive and refined exploration of the complex 3D design space.

In light of the above, we propose GenPara, an interactive 3D design editing system that leverages text-conditional shape parameters of 3D designs and infers the region of interest (ROI). We utilized the fine-tuned large language model (LLM) to extract text-conditional shape parameters, enabling a nuanced interpretation of text descriptions and aligning them with design parameters. Demonstrating the potential of LLMs to support design generation \cite{makatura2023can}, we extend this capability by fine-tuning LLMs to interpret complex 3D design parameters and interactively guide users through the design space.

Using 3D GenAI models like SPAGHETTI \cite{hertz2022spaghetti} and SALAD \cite{koo2023salad}, we constructed a design exploration space. By applying uniform manifold approximation and projection (UMAP) \cite{mcinnes2018umap}, this space was reduced to two dimensions, simplifying the visualization of complex design elements and supporting designers in navigating the 3D design process. A Bayesian inference \cite{box2011bayesian} defines the ROI of each designer by generating design alternatives through the LLM. Furthermore, we introduced a Design Versioning Tree, a novel method for visualizing the design variation structure, supporting design exploration, and enabling designers to thoroughly comprehend and engage with shape parameters generated by the fine-tuned LLM. Qualitative evaluations confirmed the precision of fine-tuned LLMs in generating text-conditional shape parameters. Subsequently, the user study demonstrated that designers successfully understood their parameters and specified exploration goals, thereby assessing the potential of GenPara for enhancing the design process. In summary, this research contributes by:

\begin{itemize}
  \item Extracting text-conditional shape parameters by training an LLM on complex 3D design elements.
  \item Generating design alternatives that reflect the shape parameters of design elements that are difficult to describe in text.
  \item Proposing an interactive 3D design editing system that enables designers to comprehend and apply text-conditional shape parameters in the design process.
  \item Demonstrating that a text-conditional shape parameter can be efficient and effective in exploring design alternatives and specifying detailed goals.
\end{itemize}

\section{Related Works}
This research focuses on improving the design generation process with 3D GenAI by accurately extracting complex shape parameters, generating user-aligned design alternatives, and visualizing these parameters. To achieve these objectives, we conducted a comprehensive review of prior work in design generation and recommendation based on user intent, GenAI applications in 3D design, and visualization of design information.

\subsection{Design Generation and Recommendation Aligned with User Intent}
Accurately inferring and delivering designs that align with user intents are pivotal challenges in the design process. Various methodologies have been developed to infer and recommend designs that satisfy user desires. Among these, the prompt engineering approach, which utilizes LLMs based on user text inputs to generate designs, is widely used. For example, systems such as GPT-4 \cite{achiam2023gpt} and DALL-E \cite{ramesh2022hierarchical} generate creative outputs from textual inputs provided by users, highlighting the potential of LLMs in design generation. In manufacturing-focused workflows, LLMs have been shown to seamlessly integrate the design generation, optimization, and manufacturing phases, by enabling users to dynamically update their constraints \cite{makatura2023can}. Furthermore, systems such as GenQuery  \cite{son2023genquery} can effectively articulate user intent more precisely, particularly in visual searches, by leveraging LLMs to transform initial abstract queries into concrete search parameters. Directgpt \cite{masson2024directgpt} describes how to interact with LLMs through direct manipulation, where users can select the part to modify in 2D images.

Despite the advantages of utilizing text inputs to derive design outcomes, LLMs face challenges in comprehensively understanding and incorporating the complex intentions and preferences of users. While GenAI systems such as GPT and DALL-E are capable of generating creative outputs from user text inputs, they face difficulties in fully reflecting complex user intentions or detailed preferences for 3D designs. This challenge arises from the the inherent spatial and structural complexities of 3D designs. Accurately describing complex objectives and specific shapes through text alone is challenging, and the description can vary considerably from one individual to another. This limitation is especially evident in the early stages of design when ideas are ambiguous and incomplete. As a result, the effectiveness of using user inputs or text-based preference inference and design recommendations is limited. To address this issue, ongoing research aims to refine the design process by accurately analyzing user preferences and intentions using Bayesian approaches. For example, systems such as CreativeSearch \cite{son2022creativesearch} and BIGexplore \cite{son2022bigexplore} utilize probabilistic interpretations of user behavior to offer design feedback by drawing on the Bayesian information gain (BIG) framework. Kadner et al.\cite{kadner2021adaptifont} introduced “Adaptifont,” a system for generating fonts through Bayesian optimization, which considers the reading speed of the user as an objective measure.  Koyama et al. \cite{koyama2022bo} extended this by employing the Bradley-Terry-Luce (BTL) model to infer relative user preferences across diverse design contexts. This method has been applied to scenarios such as selecting specific images from multiple options \cite{koyama2020sequential} or interpreting slider manipulation data \cite{koyama2022bo}. While originally framed as a preference-optimization approach, this model offers flexibility in inferring the relationships between designs, making it broadly applicable to interactive design systems.

Building on these methodologies, our framework integrates text descriptions with 3D design parameters, enabling a nuanced interpretation of user intent. By addressing the limitations of static textual prompts in capturing complex design goals, we aim to support iterative exploration and better align with user interests.

\subsection{GenAI of 3D Design}
In 3D design, the application of GenAI is rapidly advancing, revolutionizing design generation, combination, and transformation processes. Hui et al. \cite{hui2022neural} introduced a wavelet-based diffusion network that excels at generating, manipulating, and interpolating high-quality 3D objects at the part level. Furthermore, Hao et al. \cite{hao2020dualsdf} and Hertz et al. \cite{hertz2022spaghetti} developed models that manipulate specific parts by deconstructing the implicit shapes (i.e., representing 3D objects continuously through mathematical formulas) of objects, thereby facilitating precise part-level manipulations. A notable advancement is SPAGHETTI \cite{hertz2022spaghetti}, which allows for detailed control over part-level shapes using extrinsic latent elements, representing the overall shape, and intrinsic latent elements, capturing finer details. Koo et al. \cite{koo2023salad} enhanced the output quality and differentiated the training of low-dimensional extrinsic latent data from high-dimensional intrinsic latent data through a diffusion-based network, enabling text-guided part completion. Sella et al. \cite{sella2023spic} expanded these possibilities by leveraging 3D diffusion models with auxiliary guidance shapes for 3D stylization and semantic shape editing and transforming basic abstractions into highly expressive shapes using text-conditional abstraction-to-3D. 3DALL-E \cite{liu20233dall} has introduced a Fusion 360 plugin that employs text-to-image GenAI to streamline early 3D design workflows. Beyond text-driven methods, researchers have introduced intuitive shape-editing operators within the latent space \cite{hu2024cns} and facilitated 3D part editing directly from sketches, optimizing computational and time efficiencies \cite{bandyopadhyay2023doodle,bandyopadhyay2024sketchinr}.

Systems refining 2D images often employ iterative methods that integrate conditional constraints through sketches and text \cite{sarukkai2024block,zhang2023adding}. In contrast, 3D design poses unique challenges in capturing the morphological and structural complexities of 3D objects, highlighting the necessity of part-level shape editing. Despite advancements in automation and efficiency for 3D design, there remains a notable gap in supporting the intuition and creativity of designers during the actual design process. Therefore, we introduce a new approach to understanding the relationship between 3D design and text. Accurately capturing and understanding the intentions of the designer is critical for 3D design editing systems to provide relevant support in the early design process. If GenAI can better understand shape parameters that are difficult to articulate through text, it can generate more relevant design alternatives and enhance the overall efficiency of the design process. To achieve this, we incorporated shape parameters that describe the morphological characteristics of 3D shapes, enabling the system to gain a deeper understanding of the intentions of the designer. Specifically, our approach focuses on enabling designers to interactively refine design details, such as part-level parameters.

\subsection{User Interface and Visualization for Information Exploration}
Visualizing design parameters is vital for enabling designers to intuitively understand and navigate the intricate web of these parameters. This is based on the idea that when exploring information, individuals must identify an appropriate representation to encode it, as organizing and interpreting data without adequate support can be challenging \cite{suh2023sensecape}. To address this, we reviewed literature on the design process and the presentation of complex information to users. For example, Sensecape \cite{suh2023sensecape} structures information through multilevel abstraction for complex queries generated by LLMs, enhancing user efficiency and comprehension in exploration. Additionally, there are trending studies on the hierarchical visualization of information by applying ontology concepts to structured data in an organized manner \cite{jiang2023graphologue,li2022gotreescape,zhang2021conceptscope}. Tools such as Graphologue \cite{jiang2023graphologue} offer visualization of LLM-generated responses as hierarchical graphical diagrams, aiding in information retrieval and question-answering tasks, whereas ConceptScope \cite{zhang2021conceptscope} employs bubble tree visualization to map conceptual relationships within documents, such as research papers, facilitating easier reviews. Luminate \cite{suh2024luminate} introduces a new interaction paradigm with LLMs that generates key dimensions from an initial prompt to construct a structured design space, enabling users to explore and evaluate diverse creative options systematically. Juxtaform \cite{pandey2023juxtaform} facilitates the structured exploration of large corpora, reducing visual clutter while preserving clarity. SketchSoup \cite{arora2017sketchsoup} focuses on early-stage ideation by enabling the interpolation and refinement of design sketches, while  Nicolas et al. \cite{rosset2023interactive} integrates aerodynamic feedback, allowing designers to iteratively balance performance with visual aesthetics. Although these approaches focus on 2D-based exploration, our study adapts principles similar to the more complex context of 3D design, providing enhanced interactivity and a structured framework for navigating intricate design spaces.

Hierarchical visualizations, such as treemap \cite{shneiderman1992tree}, organize relationships between elements in a structured layout, allowing designers to clearly trace iterations and effectively compare different branches within the hierarchy \cite{dudavs2018ontology}. Such structures help designers trace the evolution of their work, offering an organized approach to managing iterative design processes. Davis et al. \cite{urban2021designing} on hierarchical visualization in co-creative systems within virtual environments explores how users can visually interpret and interact with parameter relationships. Demelo and Sedig \cite{demelo2021forming} form cognitive maps of ontologies using hierarchical visualizations to assist users in navigating intricate information structures and effectively tracing dependencies and transformations. Furthermore, methods for dimensionality reduction and visualization of design information using techniques, such as UMAP and VAE, have been explored \cite{li2022gotreescape,tkachev2022metaphorical,wang2023drava}. These methods embed 2D images into the design space, allowing for the confirmation of clusters or visualization of the design process in a tree-like structure. 
 
 Building on these insights, we introduce an Exploration Map and a Design Versioning Tree. These tools aim to streamline the dimensional complexity of 3D shapes and elucidate the parameter and design relationships. By adopting these tools, designers can effectively navigate between abstract exploration and detailed parameter adjustments, fostering a more intuitive and structured 3D design process.

\section{Research Challenges and Goals}
While current 3D GenAI research predominantly focuses on the performance and accuracy of generated outcomes, this approach falls short in capturing the inherent complexity of the 3D design process. GenAI can help designers quickly explore a wide range of designs, supporting broad exploration. However, it has limitations when it comes to creating designs that fully align with user intentions \cite{tholander2023design}. The lack of methods to develop user-intended design ideas can reduce the efficiency and potential of 3D Gen AI capabilities. To address these challenges, GenPara introduced three core tools aimed at enhancing the 3D design process: \textbf{text-conditional part-level editing (T1)}, \textbf{visualization of the design space to understand the distribution of design parameters (T2)}, and \textbf{ a hierarchical visualization capturing the design evolution of generated outputs (T3)}. Each tool is tailored to address a specific challenge in the iterative 3D design process as outlined below.

\subsection{Ambiguity in Initial Design Concepts and Ideas → Enhancing 3D Design Comprehension through Shape Parameters}
The early stages of 3D design often begin with ambiguity, making it challenging for designers to clearly and intuitively grasp design complexities. \cite{kazi2017dreamsketch}. Attempts to describe design concepts and ideas that are difficult to express in text are often inaccurate, complicating concept development and decision-making in the early stages of 3D design. To date, there are no GenAI tools that effectively refine 3D design shapes during the early stages of the 3D design process. \textit{GenPara} tackles this challenge through its \textbf{text-conditional part-level shape editing (T1)}. This tool empowers designers to refine specific parts of a 3D model by editing shape parameters aligned with user intents. By bridging the gap between abstract concepts and tangible design outcomes, \textbf{T1} enhances the design process by facilitating direct interaction with part-level shape parameters. This approach supports the comprehension and exploration of design possibilities, reduces ambiguity, and improves decision-making during critical stages of refinement.

\subsection{Complexity in Concretizing and Managing Complex 3D Shapes → Reducing Cognitive Load in Specifying and Refining 3D Shapes with Text}
Concretizing and managing complex shapes in the 3D design process is a significant cognitive challenge for designers. This process requires precise model transformations and visualizations that align with the intentions of the user. However, current 3D GenAI tools do not adequately support the easier management and concretization of complex 3D models. This can lead to inaccuracies and diminished detail in designs, ultimately reducing user satisfaction with the final outcomes. Moreover, the task of grasping and concretizing these complex shapes imposes a significant cognitive burden on designers. This process is not only time-consuming and demanding but also prone to inaccuracies, particularly when complex shapes are described using text \cite{zamfirescu2023johnny,liu2022design}. \textit{GenPara} reduces this cognitive burden by combining \textbf{text-conditional part-level shape editing (T1)} and the \textbf{visualization of the design space (T2)}. \textbf{T1} enables users to make precise adjustments to specific parts of a model aligning the final output more closely with their creative goals. Meanwhile, the \textbf{T2} provides a visual overview of the design space and shows the distribution of the generated text-conditional shape parameters. This combination enables designers to manage complex shapes with greater ease and precision, ensuring that their creative intent is reflected in their final designs.

\begin{figure*} [t]
  \includegraphics[width=\textwidth]{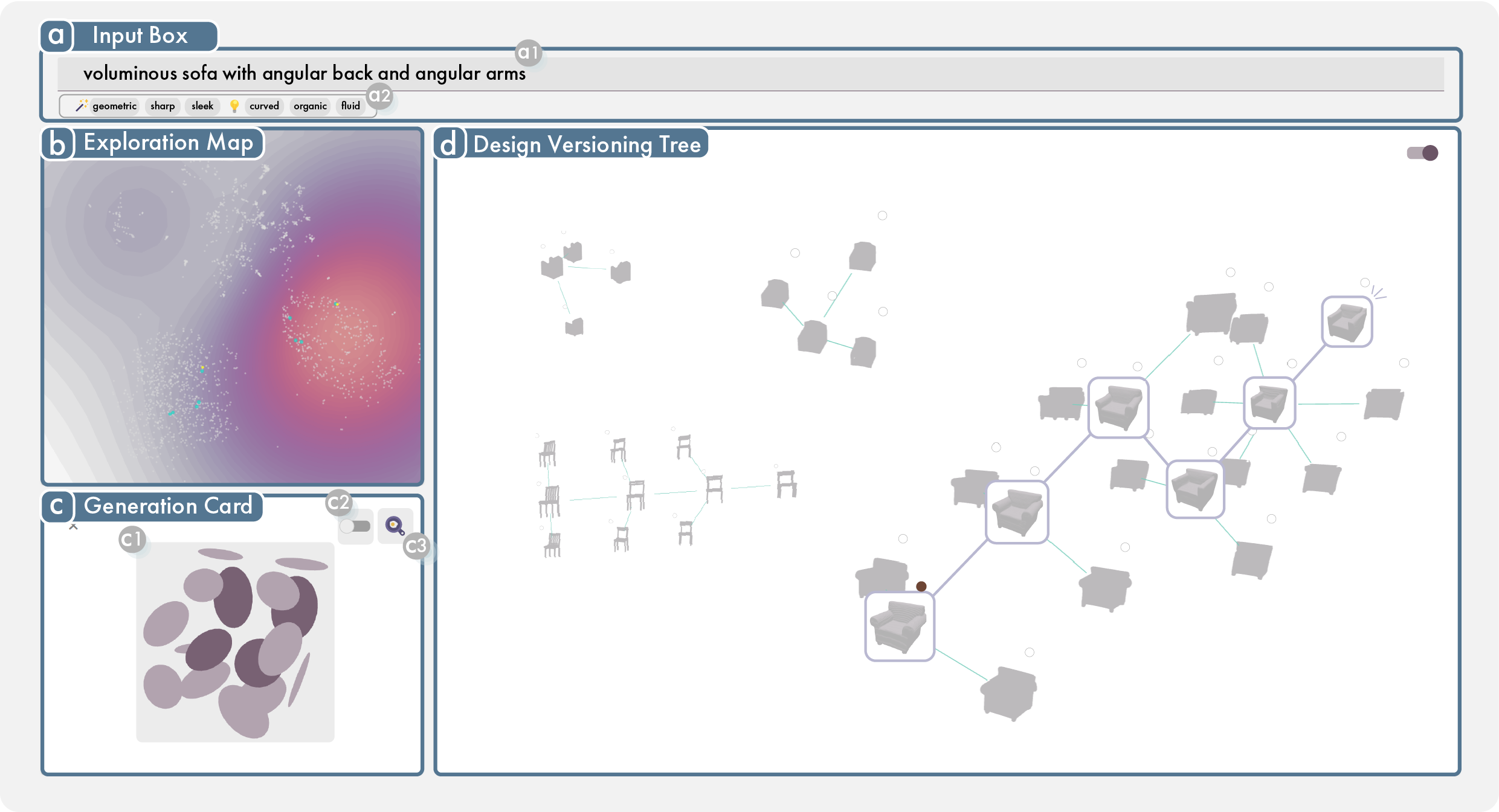}
  \caption{Overview of \textit{GenPara}: This figure demonstrates how designers utilize \textit{GenPara} to navigate and refine their design process. Designers generate initial chair concepts by providing text prompts (a1), supported by aligned and diversified adjectives (a2), in the Input Box (a). They explore potential designs on the (b) Exploration Map, represented as diverse dots. Users can select specific parts of a chair using 3D Gaussian blobs, as shown in (c1) 3D Gaussian Blob and Mesh, and switch between mesh and blob modes via (c2). The user uses (c3)  (LLM Generation Button) to create and display new design alternatives in the Design Versioning Tree on the right (d), allowing users to review and evaluate the adjustments made to the chair design. This integrated approach supports a systematic exploration from conceptualization to detailed specification.}
  \Description{Screenshot of the GenPara system interface, showcasing the process of 3D chair design. The interface is divided into distinct sections: the Input Box where initial chair concepts are typed, the Exploration Map displaying potential chair designs as colorful clusters, a Generation Card for selecting and modifying chair parts, and the Design Versioning Tree on the right illustrating the hierarchical relationships among various chair designs.}
  \label{fig:2}
\end{figure*}

\subsection{Challenges in Thoroughly Comparing and Evaluating Various Design Alternatives → Enhancing Efficiency and Creativity in Design Process.}
The design process inherently involves exploring various design options and making optimal choices. However, the complexity of design tasks and cognitive limitations pose significant challenges in exploring all possible design spaces and continuously making informed decisions \cite{kwon2024comparing,xu2023augmenting}. The repetitive processes of sketching, conceptual design, and 3D modeling further exacerbate these challenges, can diminish the overall efficiency and creativity of the design process \cite{liu2022design}. \textit{GenPara} overcomes these challenges by incorporating the \textbf{visualization of the design space (T2)} and the \textbf{hierarchical visualization capturing the design evolution of generated outputs (T3)}. \textbf{T2} helps designers to identify and explore designs that align with their ROI-designs. \textbf{T3} organizes the design history into a hierarchical structure, showing parent-child relationships between the original designs and their modifications. This allows designers to trace their decision-making processes, revisit earlier iterations, and compare alternatives systematically.

\section{GenPara}
\subsection{Conceptual Overview}
We present \textit{GenPara}, an interactive 3D design editing system that harnesses LLM to refine text-conditional shape parameters to aid users in their design process. \textit{GenPara} was developed for 3D design exploration and concretization, utilizing a fine-tuned LLM, an Exploration Map, and a Design Versioning Tree (Figure \ref{fig:teaser}). In the following sections, we discuss the outline of an envisioned user scenario to demonstrate its practical application, explain the various features of the system, and provide technical details of \textit{GenPara}.

\subsection{Envisioned User Scenario}

To illustrate the design and features of \textit{GenPara}, we explain how Noah uses it to design a custom chair for a client’s living room. Noah’s task is to design a chair that meets the client’s request for a unique aesthetic while balancing functional requirements for comfort and space efficiency.

Noah initially prompts for a "sofa with angular back" in the \textbf{Input Box} (Figure \ref{fig:2}a-1), and the \textbf{Exploration Map} (Figure \ref{fig:2}b) displays five chair designs as mint-colored dots. Noah receives three adjectives for aligned and diversified queries: the aligned queries are \textit{fluid, voluminous, and curved}, whereas the diversified queries are \textit{geometric, sleek, and angular} (Figure \ref{fig:2}a-2). \textbf{Exploration Map (T2)} categorizes these results, ensuring Noah can explore diverse designs while balancing similarity and diversity. This feature allows Noah to visualize the design space efficiently, enabling creative exploration while maintaining relevance to his design goals. Dissatisfied with the initial search results, Noah adds voluminous and angular from the queries to refine his search to "voluminous sofa with angular back and arms" and reviews the results. Noah finds a sofa he likes but is not satisfied with the shape transition between the back and seat and ponders how to edit it. Using \textbf{3D Gaussian blob selection (T1)}, Noah isolates and targets the specific part of the model—the back and seat transition—without altering other parts of the sofa. This text-conditional part-level editing \textbf{(T1)} enables precise adjustments that align with his design intent, avoiding the need for broad, untargeted modifications. To address this issue, he selects a Gaussian blob corresponding to the sofa’s back and seat and presses the \includegraphics[width=1em]{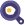} \textbf{LLM Generation button (T1)} (Figure \ref{fig:2}c-3) to explore the modification options. Shortly thereafter, three design alternatives for a more curved, organic, and faceted backseat are generated. He chooses a faceted sofa and repeats the process with the Gaussian blobs for the legs, finely adjusting the detailed goals for each part of the chair. This iterative process demonstrates how the system enables Noah to explore and refine design elements while staying responsive to specific client requirements. Finally, Noah reviews the \textbf{Design Versioning Tree (T3)} (Figure \ref{fig:2}d), which visually documents the hierarchical relationships between the generated and modified designs. T3 captures the design concretization by linking individual edits within a structured visualization. In this scenario, the Design Versioning Tree is crucial for solving Noah’s problem of tracking iterative design changes while ensuring all modifications align with the client’s requirements. By organizing designs hierarchically, \textbf{T3} allows Noah to identify dependencies between edits and evaluate the impact of specific changes on the overall design. \textbf{T3} captures the design concretization by linking individual edits within a structured visualization. For example, when Noah experiments with multiple shape transitions for the back and seat, the Design Versioning Tree provides a clear comparison of these variations, helping him justify why a specific transition was selected based on both aesthetics and functionality. This enables Noah to trace the development of his design, understand how specific edits contribute to the final result, and organize his design directions for presentation. Noah uses the Design Versioning Tree to highlight key changes and justify design decisions during a client presentation, improving communication and transparency. 

\subsection{User Interface}
\textit{GenPara} is not only capable of enabling shape generation in 3D designs that are difficult to articulate verbally but also serves as an interface that helps users understand their parameter sequences and specify their exploration goals. This system comprises the following four detailed features (Figure \ref{fig:2}): an Input Box for prompts for generating 3D chair models, an Exploration Map that enables users to explore chairs and infer the ROI, a Generation Card for selecting 3D Gaussians and generating design alternatives with LLM, and a Design Versioning Tree for visualizing the sequence of shape parameter variation.

\subsubsection{\textbf{Input Box – Prompting for Generating 3D Models}} 
\begin{figure}[h]
  \includegraphics[width=\columnwidth]{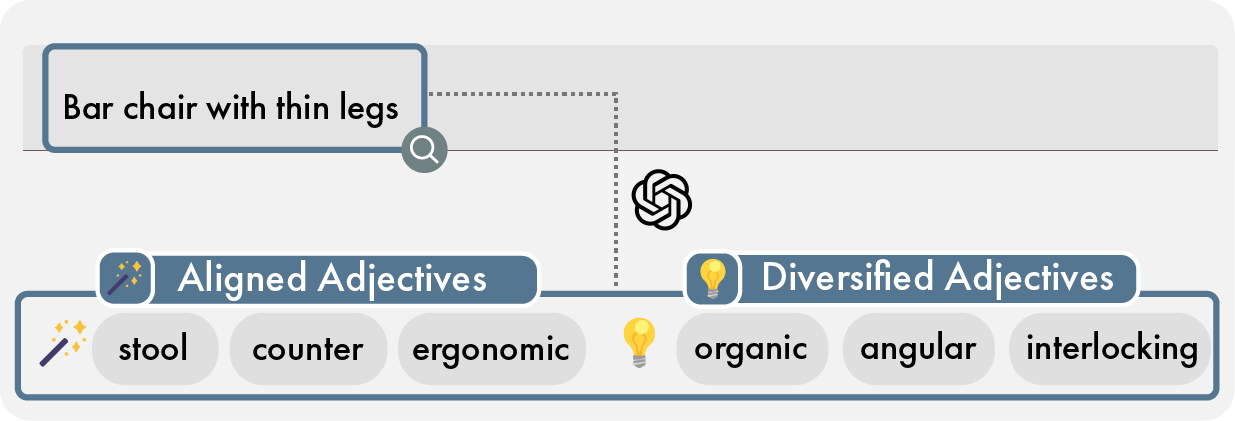}
  \caption{\textbf{Input Box}: Users can generate chair designs via text by \textbf{Input Box}. The system generates adjective suggestions based on the prompt history of the user, such as “ergonomic,” “organic,” and “angular.” These suggestions enhance creative exploration by enabling users to effectively refine their design prompts, fostering the creation of customized 3D chair models that meet specific aesthetic and functional requirements.}
  \Description{The Input Box feature of the GenPara system, showing a search bar with the typed prompt 'Bar chair with thin legs'. To the right, two tabs labeled 'Aligned Adjectives' and 'Diversified Adjectives' display suggestions such as 'stool', 'counter', 'ergonomic' for aligned and 'organic', 'angular', 'interlocking' for diversified, enhancing prompt refinement for designing customized 3D chair models.}
  \label{fig:3}
\end{figure}

\textit{(Figures \ref{fig:2}a and \ref{fig:3})}

When users generate a 3D chair model in \textit{GenPara} through prompts, the system utilizes their text search history to update \includegraphics[width=1em]{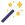} (aligned) and \includegraphics[width=1em]{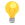} (diversified) queries with three adjective suggestions each. We developed a system that uses LLM to suggest aligned and diversified adjectives based on the history of user prompts and history of extrinsic latents of relatively ROI-designs. The method for generating aligned and diversified adjectives was based on the study of Son et al. \cite{son2023genquery} which details in Section \ref{sec:4.4.2} \& Appendix A.3. This enables support for prompting during the design process and infers part-specific adjectives for the extrinsic latents to be generated by LLMs. Consequently, based on the parameter sequence explored by the designer, LLMs can infer aligned and diversified adjectives for the shape. This ultimately allows the generation of 3D designs that incorporate modifications to the relatively ROI-designs by the user. By clicking the buttons with these adjectives, they are automatically added to the prompt. Users can then further refine and complete the prompt to generate the ROI-chair model.

\begin{figure*}[t]
  \includegraphics[width=\textwidth]{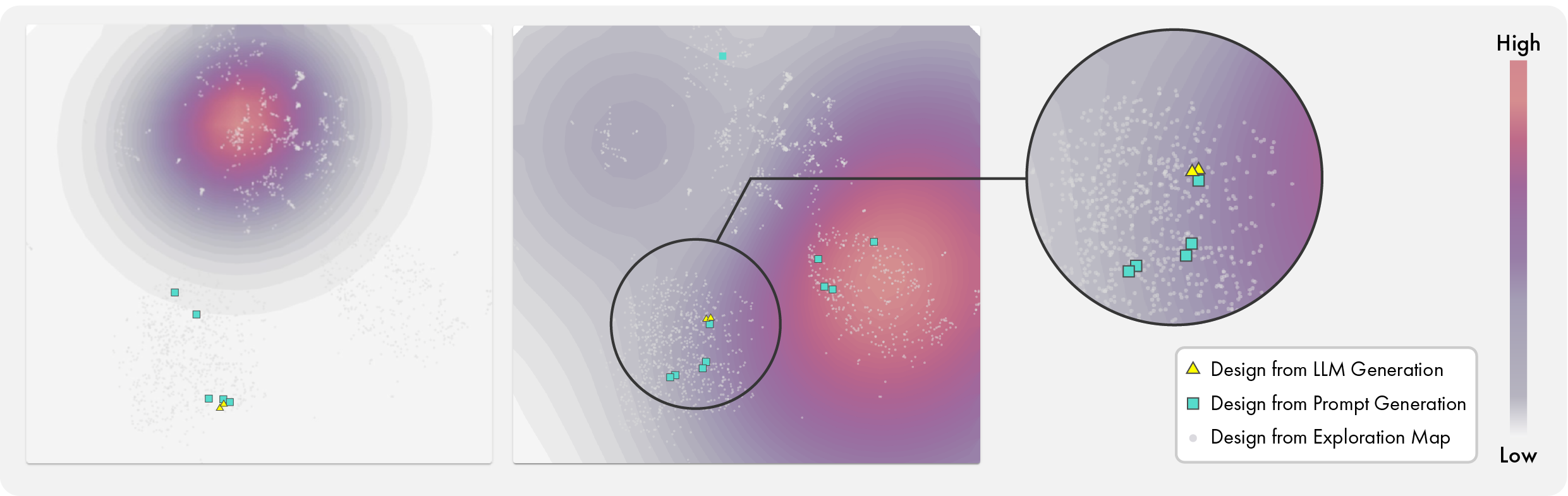}
  \caption{\textbf{Exploration Map} for the design space: This map visualizes chair designs, aiding efficient navigation through the design space. Designs generated by user prompts are marked in mint-quadrangle, whereas those from the LLM appear in yellow-triangle, enhancing user choice and exploration. As the user generates designs, the map indicates the ROI of each user in shades of pink, purple, and gray. To improve accessibility and ensure clear differentiation, the map in this paper is presented with shapes rather than solely relying on color, differing slightly from the UI used in the user study.}
  \Description{Visualization of the Exploration Map in the GenPara system, depicting clustered dots representing chair designs. The map shows user-generated designs in mint-quadrangle and LLM-generated designs in yellow-triangle. Gradient regions in shades of pink, purple, and gray indicate the ROI of each user, aiding navigation through the design space.}
  \label{fig:4}
\end{figure*}

\subsubsection{\textbf{Exploration Map and Inference ROI}} \textit{(Figures \ref{fig:2}b and \ref{fig:4})}

\begin{figure}[!]
  \includegraphics[width=\columnwidth]{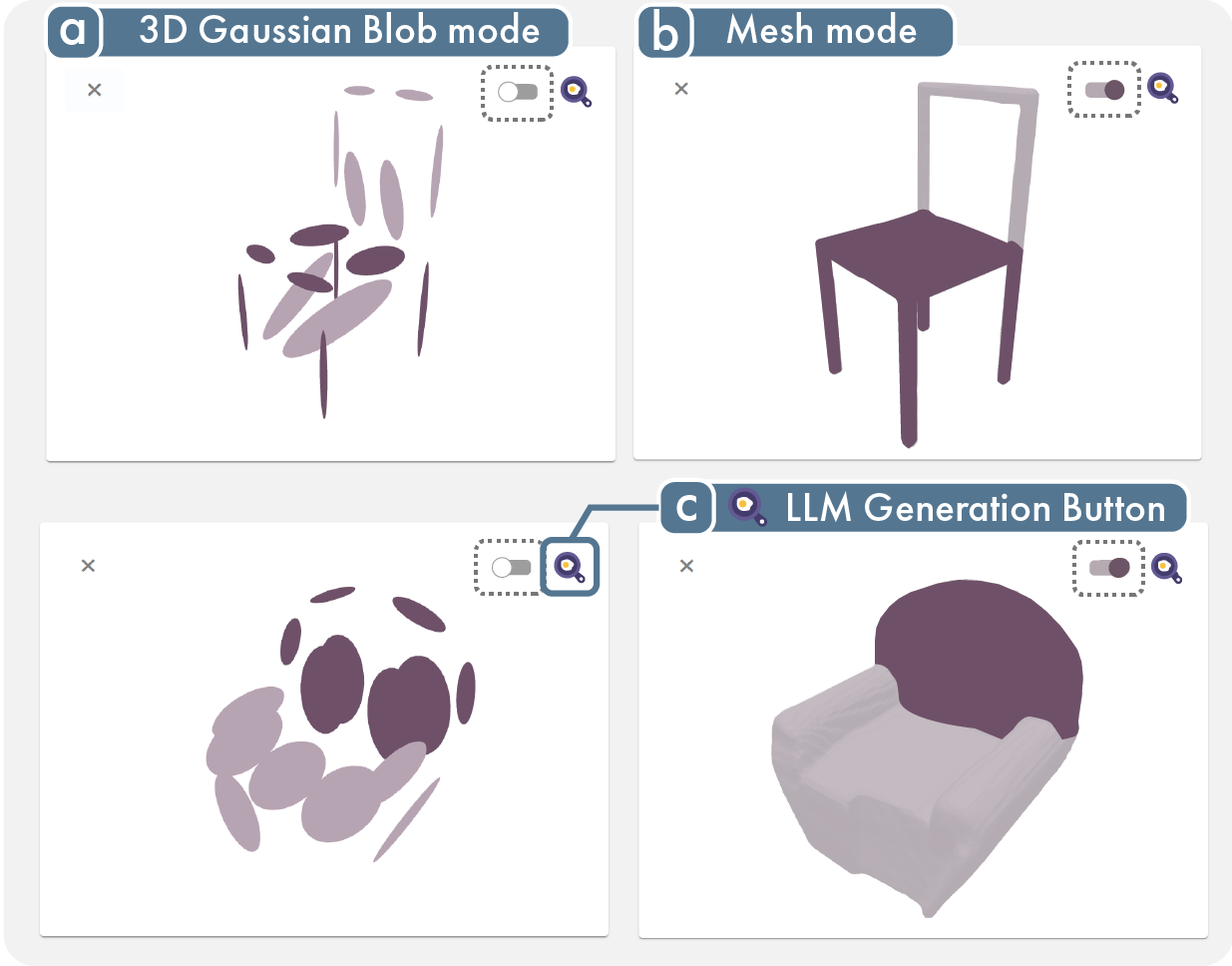}
  \caption{\textbf{Generation Card}: Users can directly interact with the 3D chair model to specify modifications. (a): 3D Gaussian blob view of the chair allows users to interactively select parts they wish to edit. Users can rotate and examine these parts closely, enhancing understanding and precision in design customization; (b): 3D mesh view of the chair also shows the selected parts as it allows users to check parts they wish to edit. Users can toggle between 3D Gaussian blob mode for an abstract representation or mesh mode for a detailed view. Selected parts are highlighted in both modes to ensure clarity and continuity in the selection process. (c) LLM Generation Button: After making their selections, users press \includegraphics[width=1em]{Figure/generationButton.png} to generate the modified design alternatives. These are subsequently displayed in the Design Versioning Tree}, providing immediate visual feedback on the changes.
  \Description{The Generation Card in GenPara displays two interaction modes. The first mode shows the chair in 3D Gaussian Blob mode for abstract part selection. The second mode provides a Mesh mode for a detailed view of the chair. Both mode has a LLM Generation Button used to create new design alternatives from selected chair parts.}
  \label{fig:5}
\end{figure}

The Exploration Map \textbf{(T2)} is a critical feature of the \textit{GenPara} system, designed to assist users in visually navigating approximately 2,000 chair designs. Each design is represented as a point on the map. Designs generated from user prompts are colored in mint and those from the suggestions of the model via LLMs highlighted in yellow. As users continue to interact with the system and generate additional designs through prompts or LLM suggestions, the number of available designs can expand beyond the initial approximate 2,000. This helps users easily distinguish between their initial ideas and generated alternatives, facilitating a more informed design process. 

Hovering over any point on the map reveals a 3D Gaussian blob of the chair, providing a preliminary visual cue for the shape of the chair. Clicking on a specific point displays a 3D Gaussian blob and mesh on the Generation Card (Figures \ref{fig:2}c and \ref{fig:5}). This seamless transition from exploration to specific design manipulation enhances user interaction by allowing quick iterations and modifications. 
Furthermore, the system employs Bayesian inference to identify ROI among LLM-generated design alternatives.  It interpreted these LLM-generated design selections as indicators of a relatively ROI-design. Regions with high calculated goodness values based
on the LLM-generated design selection are highlighted using gradient contours (Details in Section 4.4.4).

An Exploration Map facilitates user navigation through a complex array of design options by visually representing user ROI. Utilizing Bayesian inference to analyze and display ROI offers a personalized experience by highlighting regions with gradient contours. This intuitive functionality is designed to enhance the design process, allowing users to efficiently pinpoint and refine their design choices, thereby streamlining the overall workflow and leading to satisfactory design outcomes.

\begin{figure*}[!]
  \includegraphics[width=\textwidth]{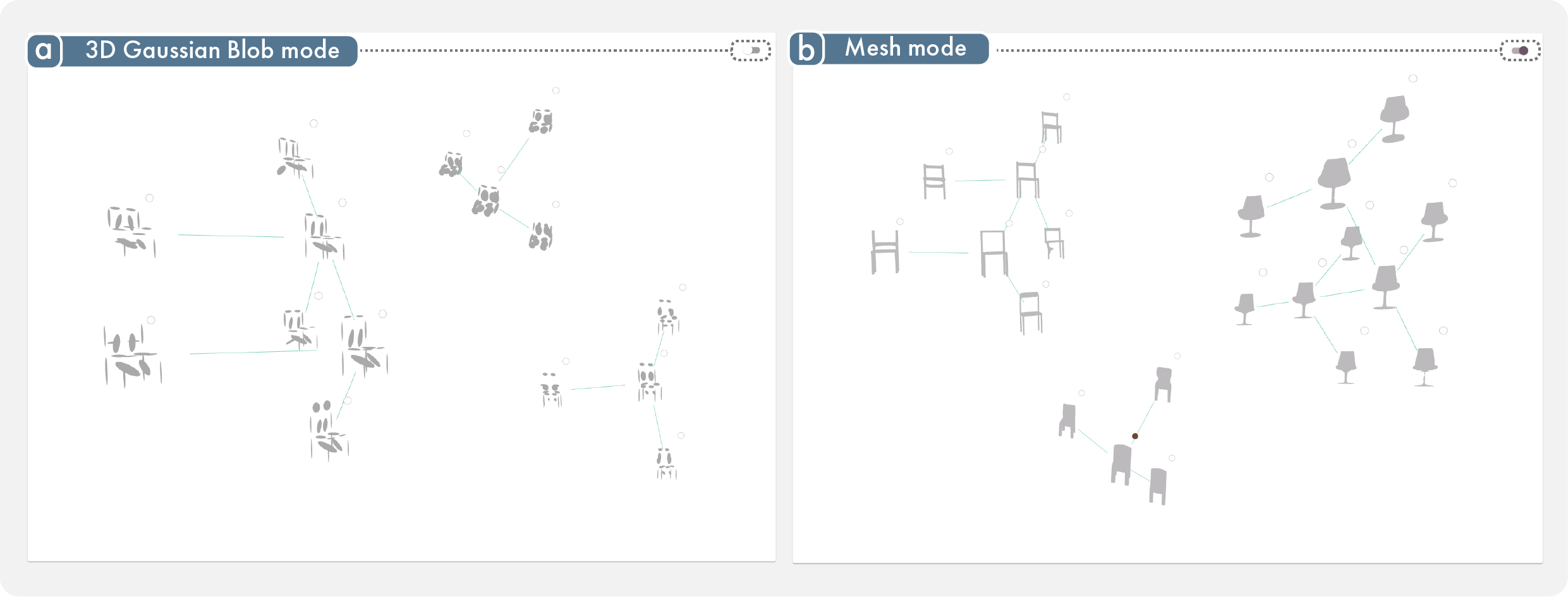}
  \caption{\textbf{Design Versioning Tree}: (a) 3D Gaussian mode, (b) Mesh mode; This figure displays a 3D node–link diagram visualizing tree-structured data of 3D design models generated by LLMs. Nodes represent individual design models while edges depict parent/child relationships organized along the y-axis to intuitively showcase the hierarchy of modified designs by users, facilitating an understanding of their focus in the design process.}
  \Description{Two views of the Design Versioning Tree in GenPara: the first in Gaussian mode showing a network of abstract chair parts as blobs, and the second in Mesh mode showing the chairs in more detail, both organizing the designs hierarchically along the y-axis.}
  \label{fig:6}
\end{figure*}

When users generate a 3D chair model in \textit{GenPara} through prompts, the system utilizes their text search history to update \includegraphics[width=1em]{Figure/alignedAdj.png} (aligned) and \includegraphics[width=1em]{Figure/diverseAdj.png} (diversified) queries with three adjective suggestions each. We developed a system that uses LLM to suggest aligned and diversified adjectives based on the history of user prompts and history of extrinsic latents of relatively ROI-designs. The method for generating aligned and diversified adjectives was based on the study of Son et al. \cite{son2023genquery} which details in Section \ref{sec:4.4.2} \& Appendix A.3. This enables support for prompting during the design process and infers part-specific adjectives for the extrinsic latents to be generated by LLMs. Consequently, based on the parameter sequence explored by the designer, LLMs can infer aligned and diversified adjectives for the shape. This ultimately allows the generation of 3D designs that incorporate modifications to the relatively ROI-designs by the user. By clicking the buttons with these adjectives, they are automatically added to the prompt. Users can then further refine and complete the prompt to generate the ROI-chair model.

\subsubsection{\textbf{Generation Card – Select 3D Gaussian Blob and Generate Design Alternatives}} \textit{(Figures \ref{fig:2}c and \ref{fig:5})}

The Generation Card facilitates interactive design modification directly within the Exploration Map. When a user clicks on a point representing their ROI-chair design in Exploration Map, the system displays the 3D Gaussian blob (Figure \ref{fig:5}a) and mesh of the chair (Figure \ref{fig:5}b). This feature allows users to identify and select specific parts that they wish to edit visually. The selected parts are highlighted in a distinct color, providing clear visual feedback and enhancing the ability of the user to precisely refine design elements. The interface allows users to switch between Gaussian blob and mesh visualization modes. When the user presses the \includegraphics[width=1em]{Figure/generationButton.png} (LLM Generation Button - \textbf{T1}) (Figure \ref{fig:5}c)., the system references the design history of the user's ROI-chairs to extract five adjectives (three-aligned, two-diversified) that represent the design's characteristics. A fine-tuned LLM then generates corresponding chair designs, creating extrinsic latents for each. These extrinsic latents are mapped using UMAP, and the system applies Bayesian Inference to calculate the most ROI-designs. The top three designs, based on these inferred values, are displayed in the Design Versioning Tree (Figure \ref{fig:6}), allowing users to review and evaluate the adjustments made to the chair design. This process ensures that the generated alternatives align with the user’s design intent without requiring additional textual input.

The system streamlines the design iteration process by selecting and modifying specific parts to generate and assess new design alternatives, thereby providing a seamless and efficient user experience for visualizing and customizing 3D chair models.

\subsubsection{\textbf{Design Versioning Tree}} \textit{(Figures \ref{fig:2}d and \ref{fig:6})}

To support users in their design process and parameter comprehending with designs generated by LLMs, we incorporated a Design Versioning Tree \textbf{(T3)} represented as a 3D node–link diagram (Figures \ref{fig:2}d and \ref{fig:6}). This diagram visualizes the data as a tree structure , where nodes represent individual 3D design models and edges signify the parent/child relationships determined by the LLMs. The parent-child relationship is established through the sequence of edits facilitated by the fine-tuned LLM. Each user-initiated edit generates a child node that is directly linked to its parent node. When a user selects a design for editing via the LLM, the system automatically produces and displays three design variations based on the specified text-conditional shape parameters. These variations, alongside the original design, were represented in the ontology map as parent nodes connected to three child nodes. Notably, the parent/child hierarchy is oriented along the y-axis, providing an intuitive display of the parameters that users have focused on editing and specifying axis-by-axis. The edges represent the hierarchical relationships that link each parent design to its corresponding child designs. This hierarchy enables users to trace their design decisions and understand how changes in specific parameters influence subsequent designs. Unlike gallery-style approaches, which display designs independently, a Design Versioning Tree captures the hierarchical dependencies between models and their edits. This structure allows users to explore design alternatives, interpret parameter relationships, and trace the evolution of their design processes. The visual representation outlines shape parameter sequences and was examined  through a user study, ensuring its support for design exploration and understanding.

\subsection{Technical Details}
\subsubsection{Part-Aware 3D GenAI Models. }

\begin{figure}[h]
  \includegraphics[width=\columnwidth]{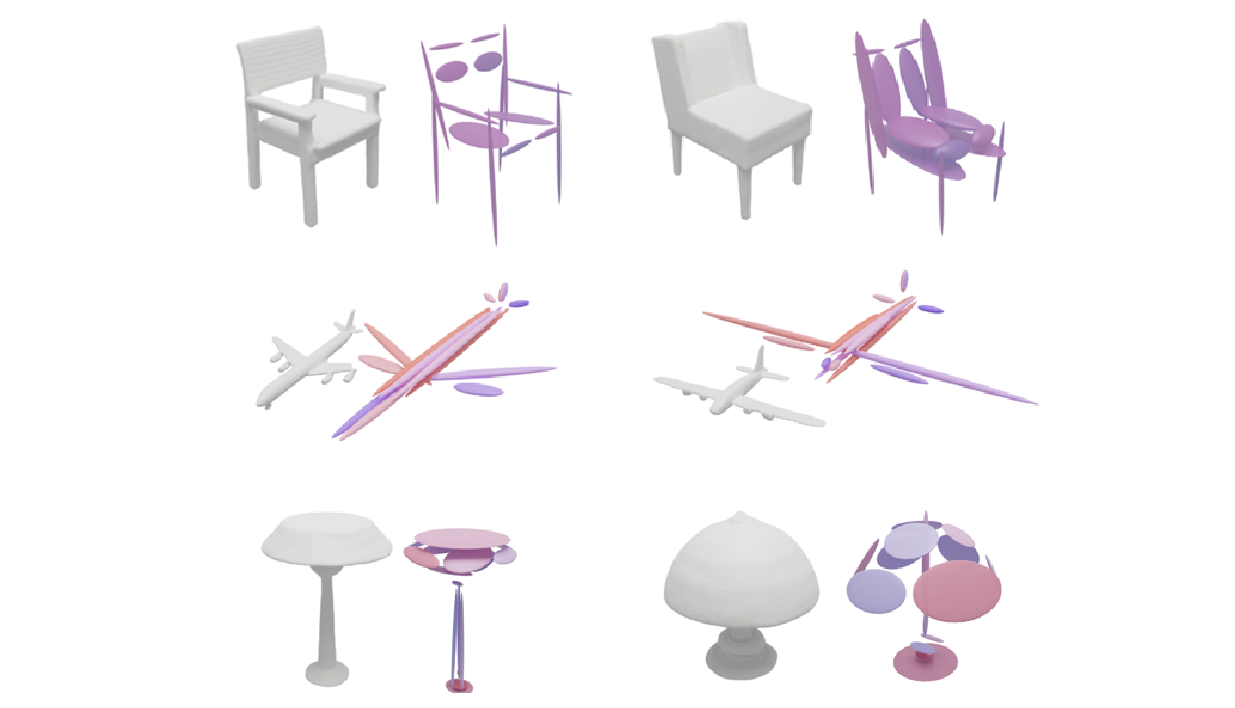}
  \caption{Gaussian blobs representing part-aware 3D models reinterpreted from the study of Hertz et al. \cite{hertz2022spaghetti}.}
  \Description{Visual comparison of part-aware 3D models with Gaussian blobs overlaid on various objects like airplanes and chairs, demonstrating different design parameters such as protrusion and thickness, reinterpreted from the study by Hertz et al.}
  \label{fig:7}
\end{figure}

SALAD \cite{koo2023salad} is a part-aware 3D GenAI system that we employed alongside SPAGHETTI \cite{hertz2022spaghetti}, which served as its foundation. SPAGHETTI \cite{hertz2022spaghetti} decomposes shape embedding into smaller and meaningful \textbf{parts embeddings} through a decomposition network trained using unsupervised learning (i.e., trained without part-level labels). The decomposition network was designed to project shape embedding onto part-level 3D Gaussians and train these parts to represent the occupancy of the actual shape. Thus, each part can be visually represented (i.e., in the form of blobs), and 3D shape manipulation is possible by modifying the Gaussian parameters, as shown in Figure \ref{fig:7}. Part embeddings are transformed into contextual embeddings that incorporate information about relationships between parts through the mixing network and are subsequently reconstructed into 3D shape via the occupancy network. In this process, the occupancy network predicts whether an input query coordinate (i.e., a specific point in the 3D space; the density of these points can vary depending on the resolution setting) belongs to a shape based on contextual embeddings, ultimately generating a 3D shape.

Returning to part embeddings, in SPAGHETTI, part embeddings are further decomposed into \textit{intrinsics}, which are trained to represent surface details, and \textit{extrinsics} (\(e\)), which are trained to represent 3D Gaussians. The collection of extrinsic latents for each part \(i\) is denoted as \( e_i = \{c_i, \lambda_{i1}, \lambda_{i2}, \lambda_{i3}, u_{i1}, u_{i2}, u_{i3}, \pi_i\} \). Specifically, \( c_i \in \mathbb{R}^3 \) marks the mean of the Gaussian mixture, \( \lambda_i \in \mathbb{R} \) represents the eigenvalues of the covariance, \( u_i \in \mathbb{R}^3 \) represents the eigenvectors, and \( \pi_i \in \mathbb{R} \) denotes the blending weights, highlighting the relative importance of each part. In detail, \( c_i \) provides the location data for each blob, \( u_i \) indicates direction, \( \lambda_i \) defines scale, and \( \pi_i \) sets the blending weights for the relative importance among adjacent Gaussian blobs. Koo et al. [16] introduced SALAD, a dual-phase diffusion model that generates the extrinsic \( \{e_i\} \) and intrinsic \( \{s_i\} \) latents of SPAGHETTI and then feeds them into the occupancy network to enhance the generation of higher-level 3D shapes and to enable the generation of 3D models based on text descriptions. In our study, we focus on generation based on the manipulation of these extrinsic parameters with SALAD \cite{koo2023salad}. This methodological choice, grounded in the strength of part-level disentangled representations and the adaptability of diffusion models, provides an effective framework for achieving state-of-the-art results in 3D shape generation and manipulation tasks. 
We utilized SALAD to generate a dataset of 58,570 chair shapes, derived from large-scale chair-shape descriptions provided in  \cite{koo2023salad,mittal2022autosdf}. This dataset was used to enable UMAP to effectively embed diverse 3D shapes. Additionally, we generated part-aware 3D models and visualized them as 3D Gaussian blobs using SALAD \cite{koo2023salad} and SPAGHETTI \cite{hertz2022spaghetti}. The dataset and methodology are further detailed in the Appendix to ensure reproducibility.

\subsubsection{Fine-tuning LLMs for generating 3D Design Alternatives.}

\begin{figure*}[h]
  \includegraphics[width=\textwidth]{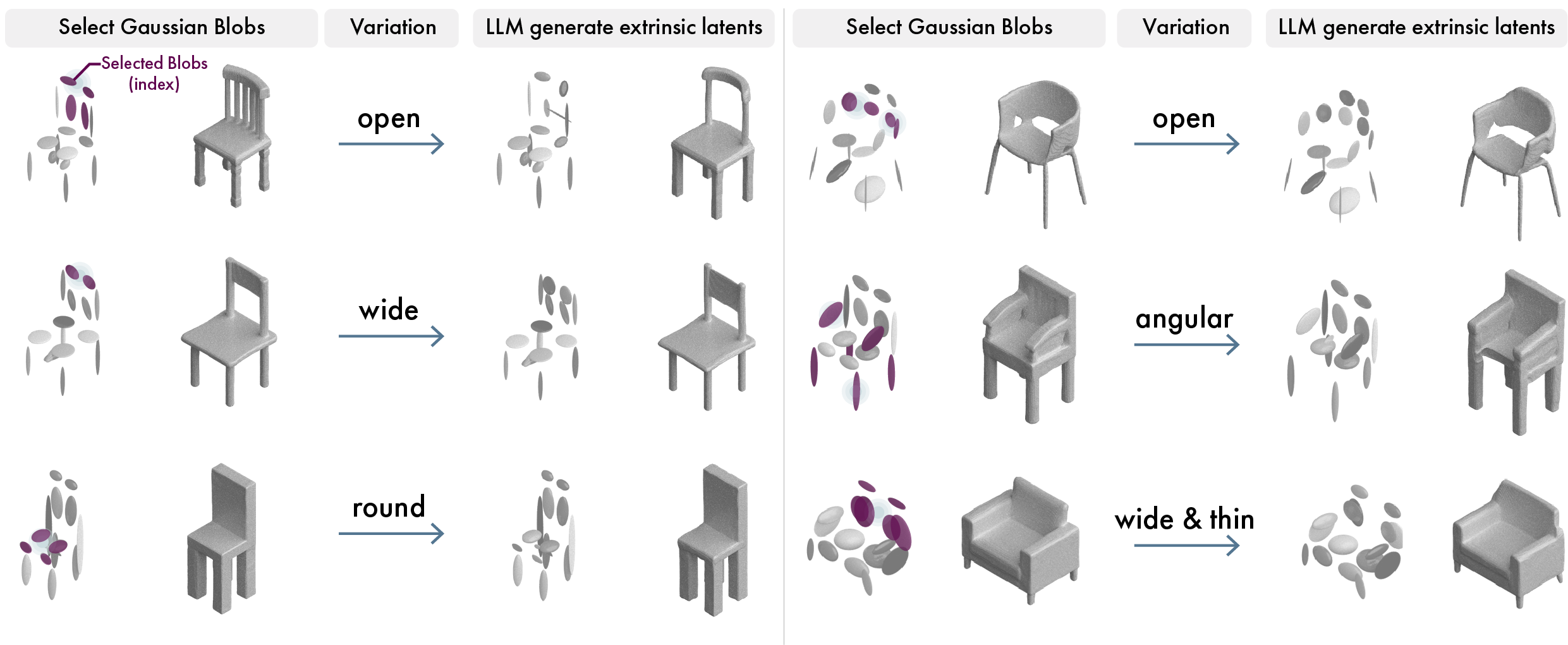}
  \caption{Qualitative evaluation of the outputs generated by fine-tuned LLM.}
  \Description{Series of images illustrating qualitative evaluations of chair designs modified by fine-tuned LLMs. Each row demonstrates a specific attribute adjustment: 'open', 'wide', 'round', 'angular', and 'wide & thin', with initial Gaussian blob representations on the left transitioning to refined chair models on the right.}
  \label{fig:8}
\end{figure*}

To effectively align 3D shape parameters with textual prompts, we fine-tuned the LLM (OpenAI-gpt-3.5-turbo-1106) using extrinsic latents generated by the 3D GenAI model SALAD \cite{koo2023salad}. For comprehensive details regarding the structure and preparation of the 3D datasets utilized in this study, please refer to Appendix A.1. Fine-tuning plays a critical role in translating ambiguous and abstract prompts into precise 3D design parameters. Without fine-tuning, adjectives like 'open' or 'curved' may result in inconsistent or irrelevant modifications. A fine-tuned model ensures these terms align contextually with specific design elements such as chair backs or armrests, enhancing the reliability and relevance of generated outputs. The 16 unique parts of a chair are represented by distinct extrinsic latents that encapsulate critical spatial information, including location, direction, scale, and blending weights. The fine-tuning process expanded these components into a sophisticated set of 256 extrinsic latents, enabling text-conditional modification of specific chair parts and robust tracking of parameter adjustments. As illustrated in Figure \ref{fig:8}, when users select a blob to edit and specify an adjective, the fine-tuned LLM provides corresponding extrinsic latent data for the 16 Gaussian blobs, facilitating intuitive and context-aware 3D design refinements.

\begin{figure*}[t]
  \includegraphics[width=\textwidth]{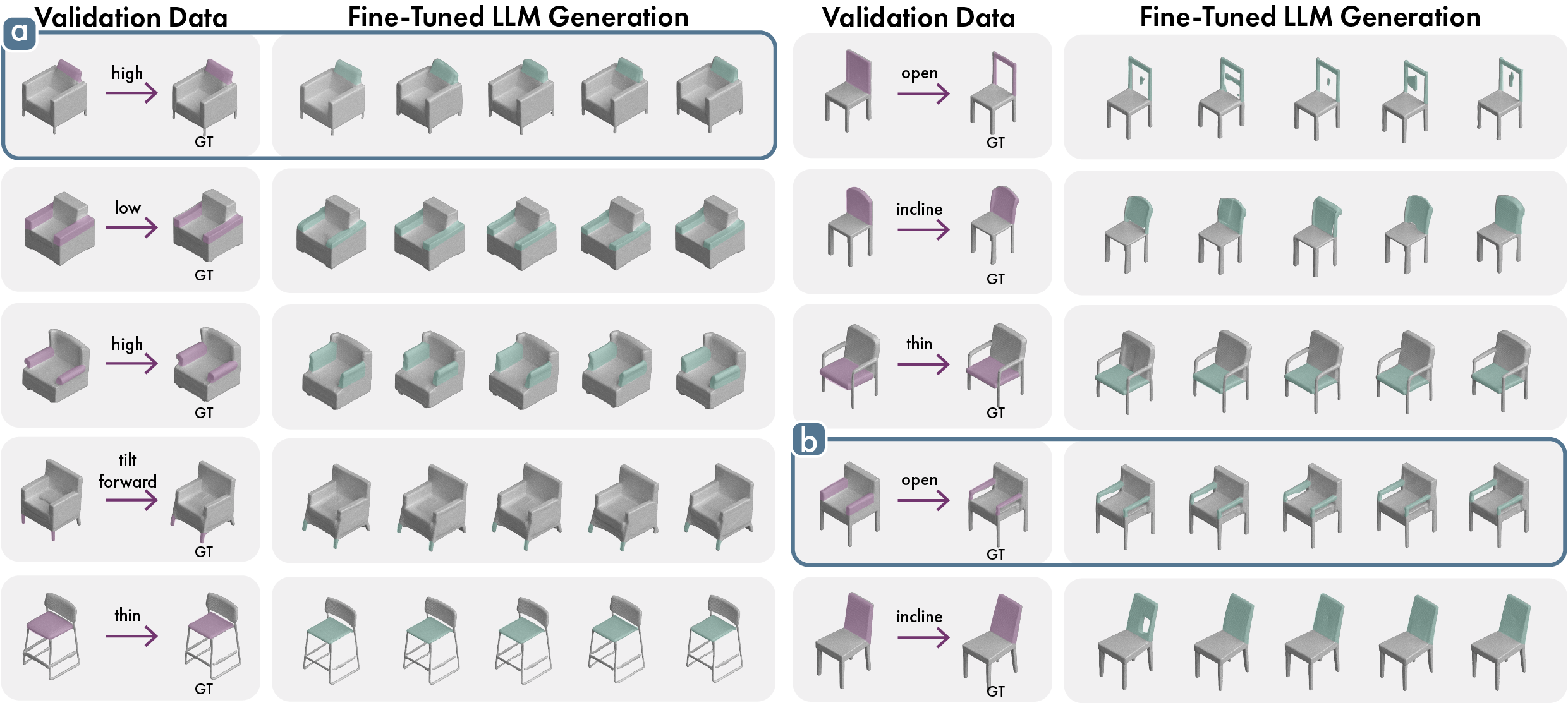}
  \caption{LLM outputs from validation datasets: (a) result where the backrest of a sofa was requested to be modified to be “high,” and (b) result of a request to make the armrests of a chair “open.” The fine-tuned LLMs successfully provided 3D models that adhered to the specified adjectives in all five attempts, demonstrating similarity to the GT.}
  \Description{Collection of 3D chair model outputs from validation datasets, illustrating LLM’s accuracy in editing chair attributes like 'high', 'low', 'tilt forward', 'thin', 'open', and 'incline' compared to ground truth models.}
  \label{fig:9}
\end{figure*}

\textbf{Training Data for Fine-tuned LLM:} As described in Appendix A.2, we used a total of 50 datasets for fine-tuning. Each dataset was designed to ensure alignment between the input prompts and expected outputs, with Gaussian-blob-based part representations serving as the foundational structure. The structure is based on the dataset format provided by OpenAI\footnote{{\href{https://platform.openai.com/docs/guides/fine-tuning}{https://platform.openai.com/docs/guides/fine-tuning}}} and is vital for training the model to generate the ROI-designs for specific types of inputs. This ensured that the model responded accurately and consistently, allowing the generation of 3D models. 

\textbf{Fine-tuned LLM Evaluation:} To evaluate how effectively our fine-tuned model adapts to new adjectives and various 3D models, we empirically designated 20\% of the training data, which equated to ten validation datasets with identical structures. From a qualitative perspective, Figures \ref{fig:8} and \ref{fig:9} demonstrate that the model successfully modified the 3D models in accordance with the adjectives provided. For instance, Figure \ref{fig:9}a shows that in all five generated outputs, the model effectively raised the backrest of the sofa, as requested, and Figure \ref{fig:9}b shows that the armrests were transformed into an open form in all trials, closely resembling the ground truth (GT) that we empirically set. The GT in these cases was determined by empirically adjusting the extrinsic parameters of the original chair designs to match specific editing goals, such as "open" or "thick." Importantly, the five generated outputs displayed in the figures represent all the attempts from the five trials, demonstrating the consistency of the performance of the model without selectively presenting the best results. The results suggested that the fine-tuned LLM can interpret user-provided adjectives and apply them to specific design parts. Qualitative evaluation indicated its potential for facilitating intuitive generation of diverse 3D model variations. While Figure \ref{fig:9} demonstrates the successful application of adjectives in most cases, some generated outputs exhibited minor deviations, reflecting the inherent challenges in fine-tuning and validating complex 3D editing. GenPara was examined through a user study focused on the chair dataset, which includes 6,778 3D models as the exclusively numerous category in \cite{chang2015shapenet}. To evaluate its applicability further, additional validation of LLM fine-tuning was performed on table and airplane models (see Appendix A.6).

\label{sec:4.4.2}
\textbf{Adjective Suggestions with Prompt and Extrinsic Latent History:} Our approach builds upon and expands on the adjective generation method in \cite{son2023genquery} by generating targeted morphological search suggestions using both prompt history and 3D extrinsic latent history. Although Genquery \cite{son2023genquery} introduced the concept of providing both aligned and diverse adjectives based on a prompt history, our method uniquely incorporates structural data from 3D extrinsic latent data to recommend adjectives that reflect the current intent of designers. This approach generates two types of search terms: aligned terms, which closely match the designer’s recent exploration path; and diversified terms, which offer alternative directions to encourage broader creative exploration. By combining a prompt history with 3D shape-specific information, our approach ensures that recommendations are both contextually relevant and morphologically precise. To illustrate the process, detailed JSON structural examples are provided in Appendices A.3 \& A.4.

\begin{figure*}[h]
  \includegraphics[width=\textwidth]{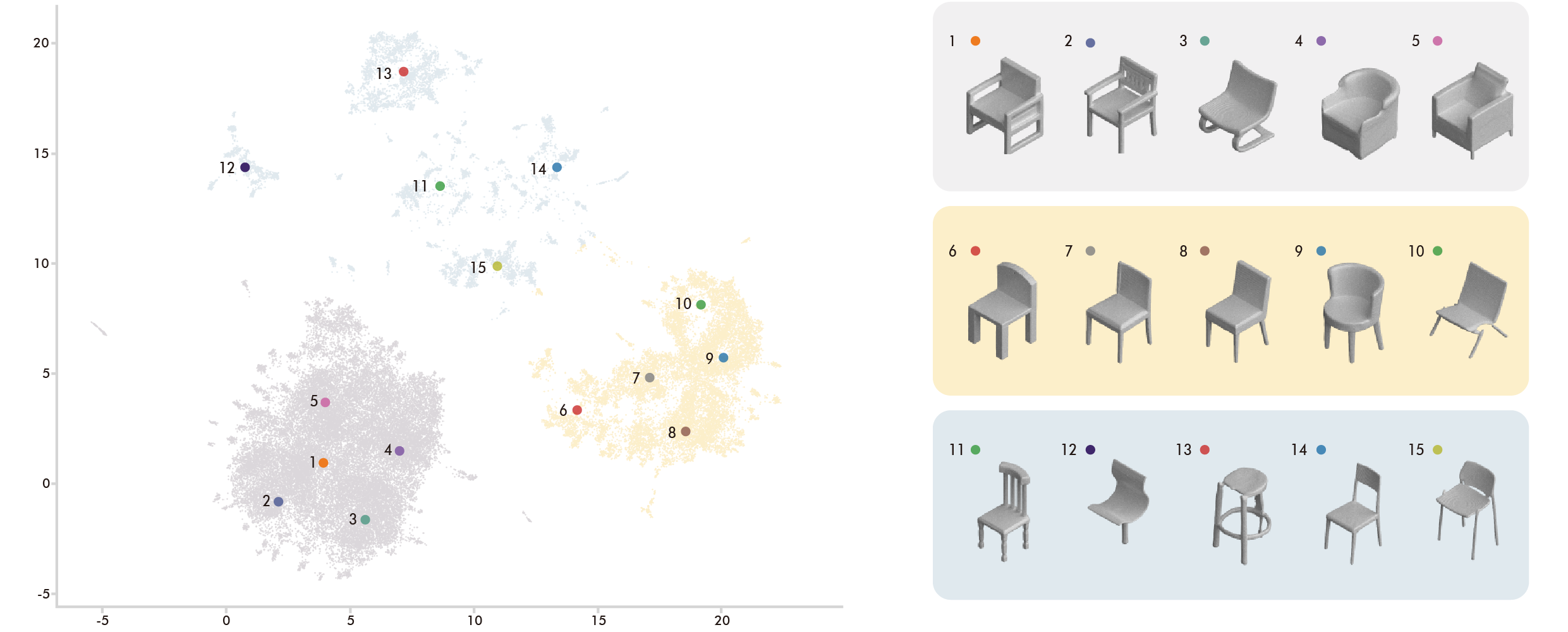}
  \caption{Distribution and clustering results of the entire dataset (3D chair models) embedded with UMAP.}
  \Description{UMAP visualization showing the distribution and clustering of 3D chair models across a scatter plot. Each cluster is color-coded to represent different design types, with corresponding chair models displayed on the right, numbered for reference.}
  \label{fig:10}
\end{figure*}

\subsubsection{UMAP for Dimension Reduction and Design Space Visualization.}

UMAP \cite{mcinnes2018umap} is highly effective for dimensionality reduction and data visualization. We embedded our collection of 58,750 3D chair models using UMAP, which excels in visualizing clusters of various design forms while preserving the structure of high-dimensional data. UMAP has several hyperparameters that significantly influence the resulting embedding, including n\_neighbors, min\_dist, n\_components, and metrics. Using the grid search method, we explored possible combinations of hyperparameters to determine the optimal set and successfully embedded the most visually similar chairs into clusters. Specifically, we set n\_neighbors = 50, min\_dist = 0.5, random\_state = 12, and metric = Euclidean to embed 58,750 chairs in 2D space using the saved model. Examining the embedded UMAP design space reveals a dense cluster at the bottom left with many armchairs or sofas, whereas the cluster at the bottom right is predominantly armless dining chairs. Additionally, various clusters at the top include stools, chairs with holes or patterns on the backrest, chairs with wheels, and an unusual number of legs. We used k-means clustering to organize the clusters into meaningful groupings (Figure \ref{fig:10}), providing a clearer understanding of the cluster organization. To balance scalability and representative diversity, we used the k-means++ algorithm to extract 1,981 representative and evenly distributed data points from the 58,750 data points. This number was selected after considering the ability to maintain design diversity across the UMAP-embedded space. Reducing the number of clusters further could oversimplify the design space, potentially excluding rare or unique design variations that are critical for comprehensive user exploration. Figure  \ref{fig:10} illustrates the distribution of clusters, highlighting key groupings such as armchairs, dining chairs, and stools. Furthermore, 3D chair models generated by SALAD or LLMs can be embedded in real time with UMAP, allowing the derivation of the parameters of each chair. This embedding capability allows the Exploration Map to dynamically reflect newly generated designs, offering users the opportunity to refine their designs interactively and review a range of available options. The Exploration Map integrates these updates, providing a structured space within the system for users to explore and navigate designs efficiently  (Figures \ref{fig:2}b and \ref{fig:10}).

\subsubsection{Bayesian Inference to Infer User-ROIs Design Parameters.}

We aim to infer the design parameters of potential ROI to users within the 2D-embedded UMAP design exploration space using Bayesian inference, specifically tracking parameters of chairs that the user selects for generation through LLMs, identifying these as relatively more interesting. To identify the ROI-designs during the design process, we utilized the method proposed by Koyama et al. \cite{koyama2022bo}, based on the Bradley-Terry-Luce (BTL) model \cite{bradley1952rank,luce2005individual}. While Koyama et al. \cite{koyama2022bo} referred to this concept as a preference, we interpreted it as interest in the context of our study to better align with our design exploration framework. This study focuses solely on modeling user-ROIs parameters using this approach, with additional mathematical formulations provided in Appendix B.

\textbf{Gradient-Based ROI Visualization in Exploration Map:} The Exploration Map uses the predicted mean of the goodness function \( g(x) \) from the Gaussian Process (GP) model to highlight areas of potential user-ROI within the 2D UMAP design exploration space. Gradients within the map visually emphasize regions with higher \( g(x) \) values, which are updated as the GP model is refined with a new design generation using the LLM. This allows users to navigate the design space intuitively, focusing on the parameters or areas of greater relevance (Figures \ref{fig:2}b and \ref{fig:4}). The process of updating the user-ROI relies on Bayesian inferences. When a user selects a design to edit through LLMs, the chosen design is of higher ROI and labeled as the “\textit{chosen option},” while the remaining designs are labeled as “\textit{other options}.” The updated GP model predicts the mean \( g(x) \) by visualizing the ROI in the UMAP space to effectively guide users through the design process. For detailed mathematical formulations, please refer to Appendix B.

\section{User Study}
To assess the impact of \textit{GenPara} on the 3D design process, we conducted a user study focused on the following research questions, examining the influence of the system on exploration and concretization of specific design objectives and design process specifications.

\begin{itemize}[label=\textbullet]
    \item \textbf{RQ1:} How does \textit{GenPara} enhance the comprehension and visualization of shape parameters in 3D design processes?
    \item \textbf{RQ2:} How does \textit{GenPara} reduce the prompting effort and facilitate specification in 3D shape design?
    \item \textbf{RQ3:} What is the perceived usefulness of \textit{GenPara} in terms of creativity and efficiency?
\end{itemize}

\subsection{Participants}
We recruited 16 participants who majored in design (age: \( M = 24.56 \), \( SD = 6.80 \); responded 7 women and 9 men). The participants, ranging from students to professionals, came from spatial design-related domains, including architecture, construction, interior design, and lighting design. Additionally, we surveyed the experience of participants with GenAI tools during the design process. Out of the 16 participants, 14 had prior experience using GenAI tools, with 12 primarily using ChatGPT for tasks such as idea generation, text refinement, and coding. Five participants had used GenAI features in Photoshop and Illustrator to concretize their design work, while the remaining participants had experience with generative image tools such as Lightroom\footnote{\url{https://lightroom.adobe.com/}}, Midjourney\footnote{\url{https://www.midjourney.com/}}, Perplexity AI\footnote{\url{https://www.perplexity.ai/}}, Claude AI\footnote{\url{https://claude.ai/}}, DALL-E\footnote{\url{https://openai.com/index/dall-e-3/}}, and PromeAI\footnote{\url{https://www.promeai.pro/}}. Two participants (P11, P16) had no experience with GenAI tools in the design process. We investigated the specific impact of \textit{GenPara} on exploration and concretization stages in the 3D design process, targeting these participants with varying levels of GenAI experience.

\begin{figure}[h]
  \includegraphics[width=\columnwidth]{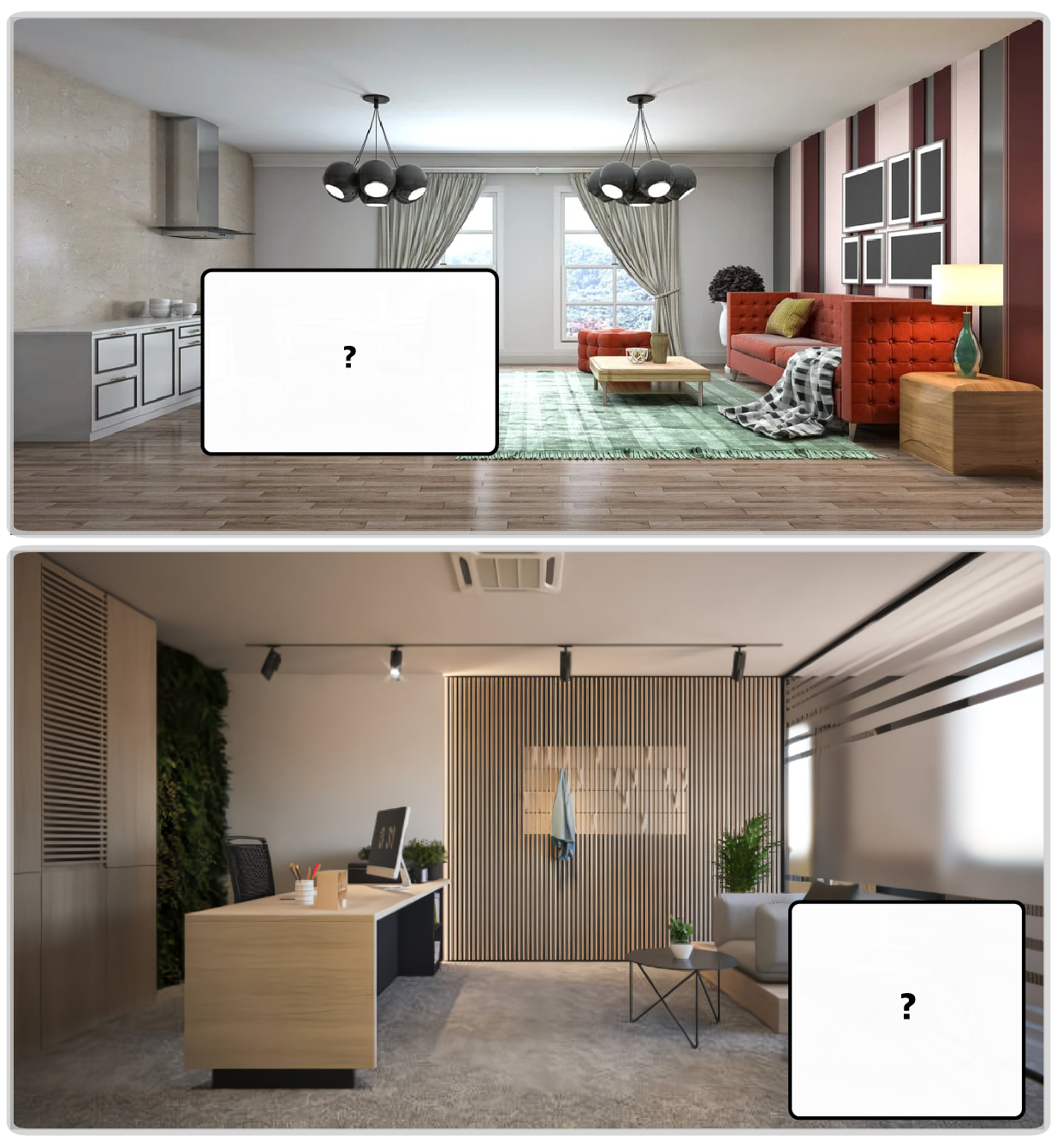}
  \caption{Two spatial contexts used in the design brief of the study. Participants were asked to explore and concretize a suitable chair to fit the white boxes for the provided interior space.}
  \Description{Two interior design contexts provided for a study's design brief. The first image shows a modern living room with space designated for a chair, marked by a white box, and the second image displays a contemporary office setting, also with a marked white box for chair placement.}
  \label{fig:11}
\end{figure}

\subsection{Setup}
All sessions were conducted offline, and we recorded the screen of all design processes and the audio of the interviews. The system used in the experiment was developed using Python 3.9. All the participants used a high-performance client server system (client: React on Windows OS with Intel Core i9 10980 XE, 64 GB RAM; server: Python Flask on Linux OS with an NVIDIA RTX 4090 graphics card).

\subsection{Procedure}
The task given in the study was to explore chair designs suitable for specific spatial contexts and to concretize the design goals in terms of shape (Figure \ref{fig:11}). Participants were informed that this task was similar to the early stages of the design process, involving exploration and design objective concretization using GenAI. The primary objective of the user study was to explore chair designs suited to the tasked interior space and fully develop the shape of one chair to complete the design. While there are various methods to generate design ideas, we specifically guided participants to use the provided search tools to perform the ideation and concretization processes using GenAI. Participants were required to finalize their 3D designs using \textit{GenPara} and a Baseline system, which has no main features of \textit{GenPara}: Exploration Map, design alternatives generated by LLM, and Design Versioning Tree (See Appendix C.1). To address the research questions outlined earlier, the Baseline system provided an equal number of 3D designs generated through prompts, which were visualized in a gallery format similar to traditional reference exploration methods like Pinterest\footnote{\href{https://co.pinterest.com/}{https://co.pinterest.com/}}, Behance\footnote{\href{https://www.behance.net/}{https://www.behance.net/}}, and Objaverse\footnote{\href{https://objaverse.allenai.org/explore}{https://objaverse.allenai.org/explore}}. Additionally, participants were instructed to concretize their design goals and the shape of their 3D designs using text. The total duration of the study was approximately two hours.

The experiment was conducted in the following detailed four stages:
\begin{itemize}
    \item \textbf{Introduction (5 minutes):} The designers were briefly acquainted with the purpose of the study and the experimental procedure.
    \item \textbf{Tutorials 1 and 2 (5 minutes each):} The participants were guided through the various features of \textit{GenPara}/Baseline and familiarized with the interface and its features.
    \item \textbf{System Interactions 1 and 2 (minimum 10 minutes, maximum 30 minutes each):} The objective was to design a suitable chair that complements the interior image (Figure \ref{fig:11}). Participants were informed that their task was to engage in the exploration and concretization phases using GenAI. After each task, participants completed a five-minute survey.
    \item \textbf{In-depth Interview (20 min):} After completing all tasks, participants discussed the difference in their experience between the two conditions in a semi-structured interview. The interview included questions on how the systems influenced their design processes, their thoughts on the provided design alternatives, and how they continued to develop their designs.
\end{itemize}

\subsection{Measures}
After each task, participants completed a survey that included questions evaluating their satisfaction with the generated designs and search process \cite{kang2021metamap}, the quality of the designs, and their behavioral intentions \cite{venkatesh2008technology} on a 7-point Likert scale. Additionally, participants assessed how well they could express their search intentions and responded to questions related to the Creativity Support Index (CSI) \cite{cherry2014quantifying}. The survey also incorporated questions from the NASA-TLX questionnaire \cite{hart1988development} to investigate any additional workload caused by the generative functions. For the comparison of survey responses between the two systems, normality was tested using the Shapiro-Wilk test. If both systems satisfied the normality assumption, a paired t-test was used; if normality was not satisfied for either system, a Wilcoxon test was conducted.

In-depth interviews focused on exploring the differences participants perceived in the design exploration and concretization processes, depending on the presence or absence of the key features of \textit{GenPara}, such as the Exploration Map, design alternatives generated by LLM, and Design Versioning Tree. Participants were also asked whether there were differences in how they concretized and expressed their design goals and 3D designs using text, and if so, how these differences impacted the design exploration and concretization processes.

We analyzed the interaction logs of participants to understand how the design process specifically progressed. Additionally, the number of prompt generations and the time spent on each design process were measured, followed by a paired t-test analysis.

\section{Results and Discussion}
The user study results indicated that \textit{GenPara} performed better than Baseline during design process. Specifically, participants discovered more diverse and creative design ideas with less effort during shape parameter-based exploration using \textit{GenPara}. Furthermore, \textit{GenPara} helped participants subdivide their design goals and proceed with the morphological concretization of each part of the 3D design, aiding in a structural understanding of the process. We observed that 3D design alternative generation by LLM helped participants better understand intricate shape parameters. These parameters, often difficult to describe in text, contributed to more satisfying design outcomes. Participants reported lower levels of frustration, higher tool transparency, and a significantly greater willingness to continue using \textit{GenPara} compared with  Baseline. The detailed discussion of the three research questions, based on the user study results, is provided in the following subsections.

\subsection{RQ1. How does \textit{GenPara} enhance the comprehension and visualization of shape parameters in 3D design processes?}
\textit{GenPara} facilitates design exploration by helping users easily understand design and shape parameters (Table \ref{Table:1}; Exploration (CSI)). Survey results showed that participants were satisfied with the creativity and variety of outcomes provided by \textit{GenPara}, indicating its effectiveness in supporting the understanding and visualization of shape parameters Figure \ref{fig:13}; Q1, $Mean_{diff}$ = 1.94, \( p < 0.001 \); Q2, $Mean_{diff}$ = 1.00, \( p < 0.010 \); Q3, $Mean_{diff}$ = 2.00, \( p < 0.010 \)). Participants highlighted that \textit{GenPara}'s intuitive visualization tools (T2 \& T3) significantly aided their understanding of the relationships between text prompts and shape parameters during the early design stages (exploration and goal concretization). The Exploration Map offered real-time insights into text-conditional shape relationships, while the Design Versioning Tree enabled tracking and refinement of shape parameters. Together, they allowed participants to manage complex parameters effectively and achieve more informed, creative design outcomes.

\begin{figure}[!]
  \centering
  \includegraphics[width=\columnwidth]{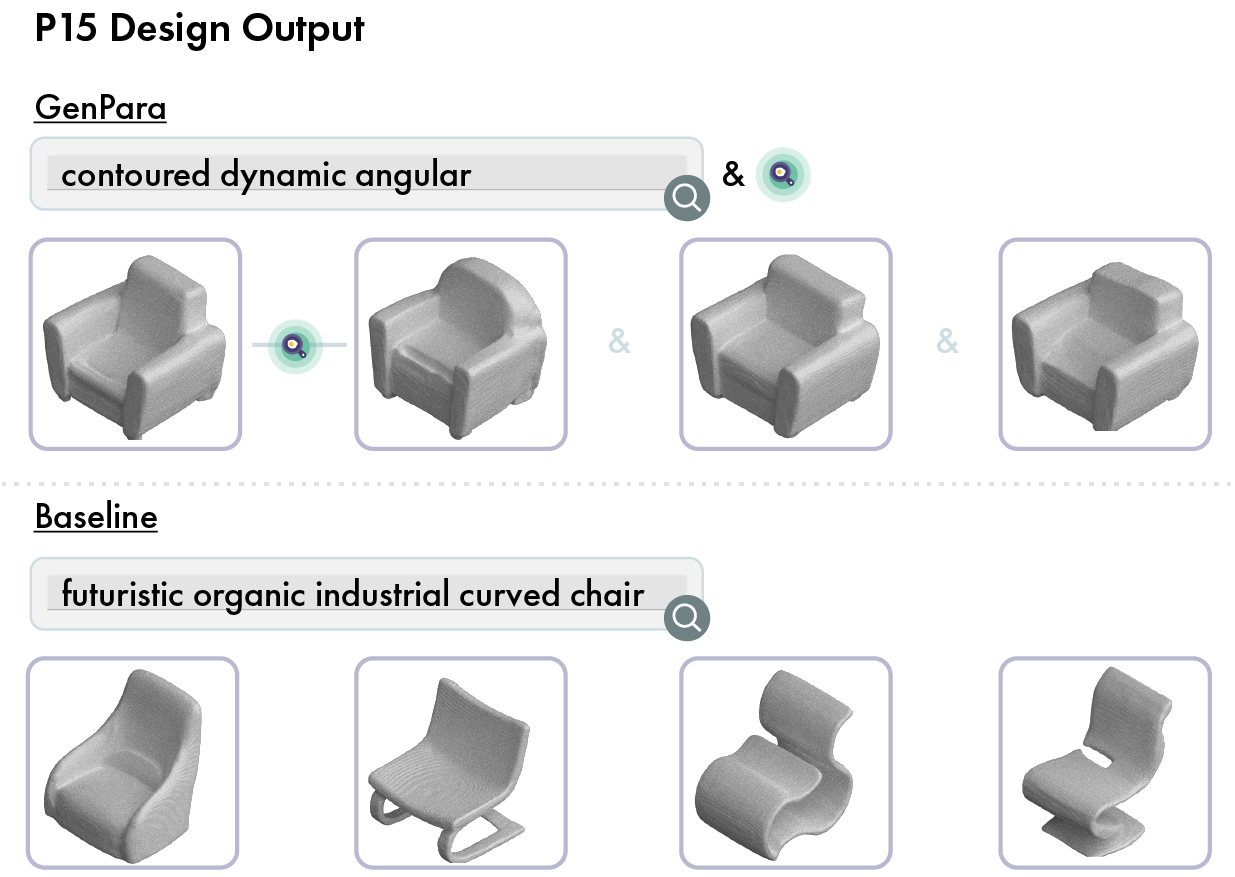}
  \caption{Comparison of design outputs generated using \textit{GenPara} and the baseline of P15.}
  \Description{Comparison of design outputs generated using \textit{GenPara} and the baseline method for Participant 15 (P15). The \textbf{GenPara} approach generated structured and incremental modifications by leveraging textual guidance such as "contoured dynamic angular," leading to a refined design exploration process. In contrast, the \textbf{Baseline} method, guided by the prompt "futuristic organic industrial curved chair," produced diverse yet less structured design variations. This visualization highlights the differences in design outcomes, showcasing how GenPara facilitates targeted and controlled refinements compared to the baseline approach.}
  \label{fig:12}
\end{figure}

\textbf{Understanding the Relationship Between Shape Parameters and Texts.} \textit{GenPara} helped participants grasp the relationship between the shape parameters and texts. Most participants noted differences in understanding the text-conditional shape parameters and their relationships when comparing the designs generated by prompts in \textit{GenPara} and Baseline (P2, P3, P5–15). P3 mentioned, “\textit{Through the Exploration Map, I could intuitively understand which design I was viewing and which area I was exploring. This contrasted with Baseline, which only presented results, and helped me easily distinguish between design types and nearby designs. It is easy to facilitate the ideation of how to concretize the design.}” P12 stated, “\textit{When generating an ‘armchair,’ the design was positioned within a similar cluster. However, when generating a ‘thin-framed chair,’ it was positioned across several clusters, making it intuitive to grasp the diversity and trends of shape parameters according to the text. This helped me quickly set my initial design goal}” (Figure \ref{fig:14}). Some participants (P5, P10, P14, P15) mentioned that while the gallery format of Baseline was convenient for viewing results at a glance, they found it challenging to understand the relationship between the prompt they had in mind and the detailed shapes of the designs (Figure \ref{fig:12}).

\textbf{Intuitive Text-conditional Shape Parameter Exploration.} \textit{GenPara} simplifies shape parameter recognition for users and expands their thoughts through shape parameter-based exploration. Most participants (P1–P13, P15, P16) mentioned that during the design concretization process, they could clearly understand the changes in shape parameters and refine their design goals. P8 stated, “\textit{The tree structure provided by the Design Versioning Tree facilitated the understanding of the relationships between text and shape parameters.}” He mentioned that when concluding his task, “\textit{Following the Design Versioning Tree along its axis allowed me to recognize the shape I wanted for my part and enabled the expansion of my thoughts.}” Additionally, P15 explained, “\textit{After quickly refining the design I initially envisioned, I used the Exploration Map to explore various design alternatives. Observing the changing colors of updating the map, I realized I was focusing too much on one area and readjusted the aspects I needed to focus on.}” Furthermore, compared to the Baseline, participants found their exploration limited to designs directly related with their prompts that is hard to compare diverse design options or identify recurring elements effectively. In contrast, the Exploration Map and Design Versioning Tree provided visual clarity and flexibility, effectively linking user inputs with design changes and supporting more precise goal refinement.

\begin{table*}[h]
  \centering
  \caption{Results of NASA-TLX (\cite{hart1988development}; 7-point Likert scale) and CSI (\cite{venkatesh2008technology}; 10-point Likert scale) survey. As \textit{GenPara} does not support collaboration with other designers, we excluded collaboration. $-$: \( p > 0.100 \), $+$: \( 0.050 < p < 0.100 \), $\ast$: \( p < 0.050 \), $\ast\ast$: \( p < 0.010 \), $\ast\ast\ast$: \( p < 0.001 \).}
  \label{Table:1}
  \Description{This table presents the results of the NASA-TLX and CSI surveys comparing \textit{GenPara} and Baseline systems. Mean and standard deviation values are shown for each survey category, along with p-values and significance levels. NASA-TLX categories include mental, physical, and temporal effort, while CSI categories include enjoyment, exploration, and expressiveness. Collaboration metrics were excluded as \textit{GenPara} does not support collaborative tasks.}
  \begin{tabular}{llcccccc}
    \toprule
    Survey  &  & \multicolumn{2}{c}{\textit{GenPara}} & \multicolumn{2}{c}{Baseline} & \multicolumn{2}{c}{Statistics} \\
    \cmidrule(lr){3-4} \cmidrule(lr){5-6} \cmidrule(lr){7-8}
    & & mean & std & mean & std & p-value & Sig. \\
    \midrule
    NASA-TLX \cite{hart1988development} & Mental & 3.31 & 1.00 & 3.88 & 1.54 & 0.16 & - \\
                      & Physical & 2.69 & 2.00 & 1.75 & 0.86 & 0.04 & * \\
                      & Temporal & 1.88 & 1.00 & 2.38 & 1.54 & 0.23 & - \\
                      & Effort & 2.94 & 1.38 & 4.25 & 1.73 & 0.02 & * \\
                      & Performance & 5.81 & 1.00 & 4.06 & 1.84 & 0.00 & *** \\
                      & Frustration & 1.75 & 1.12 & 3.06 & 1.61 & 0.03 & * \\
  \addlinespace 
  \addlinespace
    CSI \cite{venkatesh2008technology}      & Enjoyment & 8.38 & 1.70 & 5.94 & 2.49 & 0.00 & ** \\
                      & Exploration & 8.50 & 1.50 & 5.06 & 2.38 & 0.00 & *** \\
                      & Expressiveness & 6.25 & 1.39 & 3.75 & 1.53 & 0.00 & ** \\
                      & Immersion & 7.94 & 1.94 & 6.25 & 2.21 & 0.02 & * \\
                      & Results Worth Effort & 8.13 & 1.82 & 5.50 & 2.45 & 0.00 & ** \\
                      & Collaboration & - & - & - & - & - & - \\
    \bottomrule
  \end{tabular}
\end{table*}

\subsection{RQ2. How does \textit{GenPara} reduce the prompting effort and facilitate specification in 3D shape?}

\begin{figure*}[h]
  \includegraphics[width=\textwidth]{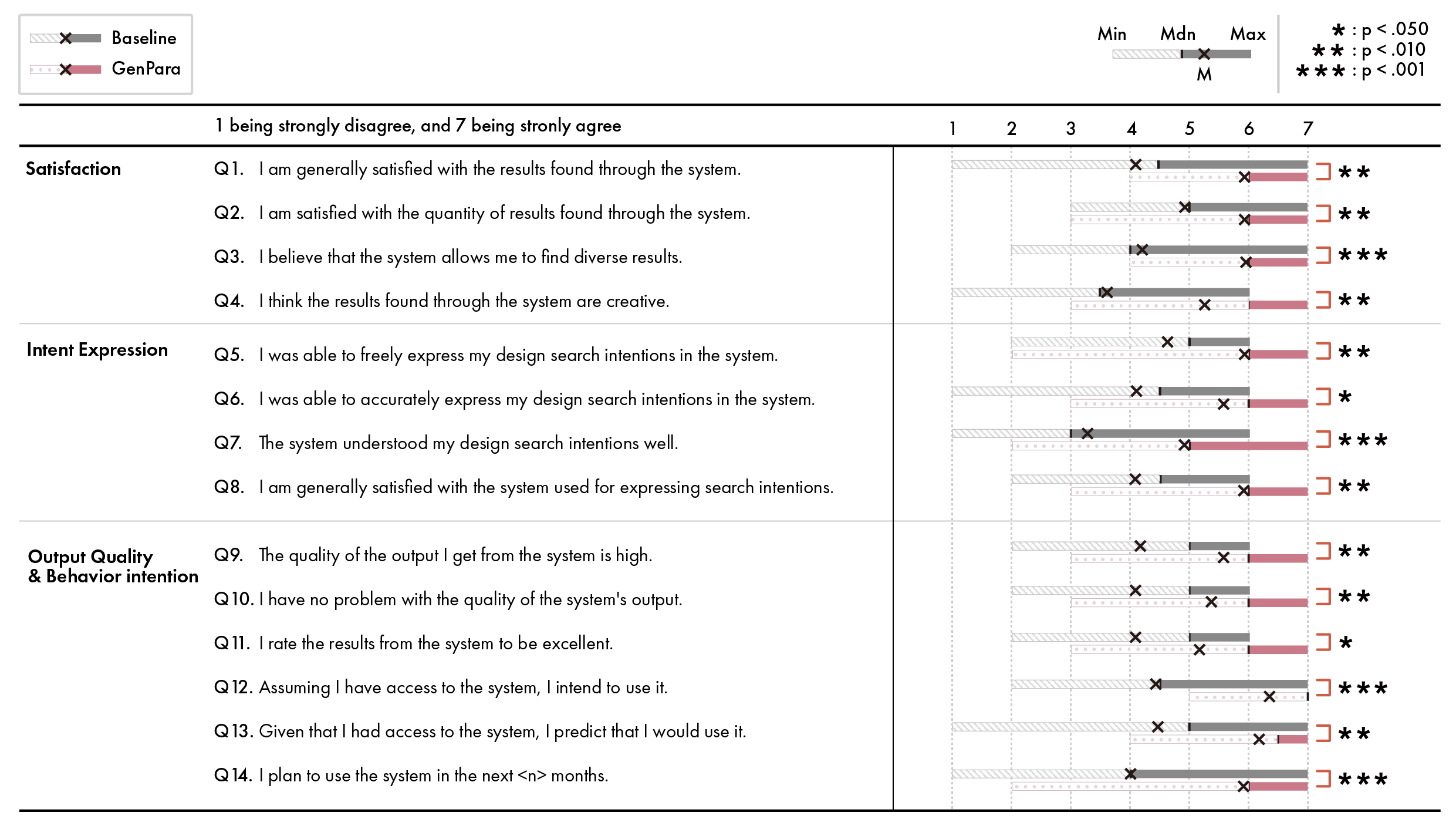}
  \caption{Survey results in the user study. The survey asked participants to rate their Satisfaction \cite{kang2021metamap}, Intent Expression, and Output Quality \& Behavior intention \cite{venkatesh2008technology} on a 7-point Likert scale. $\ast$: \( p < 0.05 \), $\ast\ast$: \( p < 0.01 \), $\ast\ast\ast$: \( p < 0.001 \).}
  \Description{Survey results bar chart comparing user satisfaction between the baseline system and GenPara across various metrics such as Satisfaction, Intent Expression, and Output Quality. Each metric is rated on a scale from 1 (strongly disagree) to 7 (strongly agree), with GenPara generally scoring higher, indicated by rightward shifts in median responses marked with statistical significance levels.}
  \label{fig:13}
\end{figure*}

\begin{figure}[h]
  \includegraphics[width=\columnwidth]{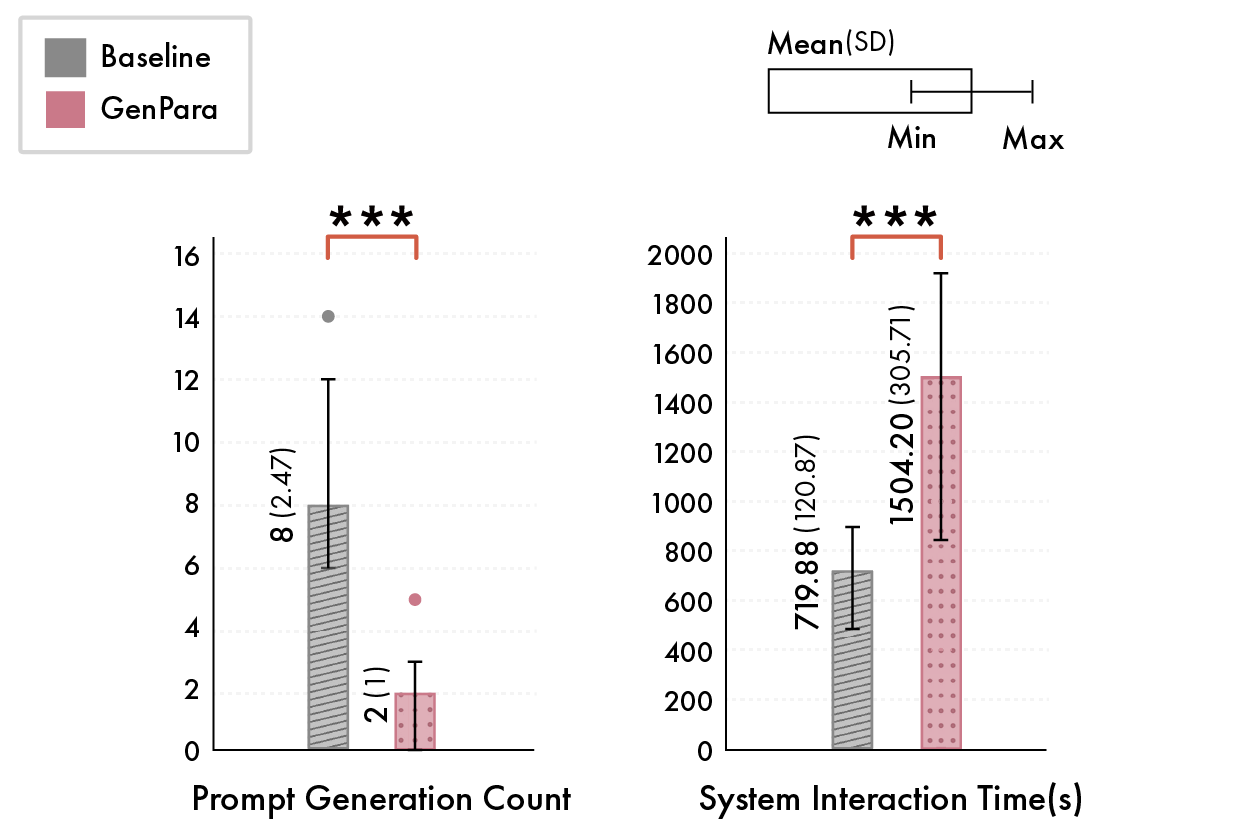}
  \caption{Comparison of system logs between \textit{GenPara} and baseline. - Prompt Generation Count (Left), Total System Interaction Time (Right). GenPara is represented by pink dotted bars, and baseline is represented by gray hatched bars.}
  \Description{Bar charts comparing baseline and GenPara in terms of Prompt Generation Count and Total System Interaction Time. The charts show GenPara resulting in higher interaction times and more prompt generations, indicating a more engaged design process.}
  \label{fig:16}
\end{figure}

\textit{GenPara} plays a crucial role in assisting users in refining detailed goals and specifying 3D design parts. \textit{GenPara} significantly reduced the burden of prompt generation by understanding and reflecting the design intentions of the user. Participants showed positive reactions to how generated design alternatives by LLM clarified shapes \textbf{(T1)} that were difficult to articulate in text and assisted in controlling detailed 3D design parts. Although the design process took longer with \textit{GenPara} (Figure \ref{fig:16}; Total System Interaction Time, $Mean_{Baseline} = 719.88$, $Mean_{GenPara} = 1504.20$, \(p < 0.001\)), participants achieved more satisfactory outcomes with less effort (Table \ref{Table:1}; Effort (NASA TLX)). The additional time spent specifying goals in detail led to greater immersion and more thorough concretization of designs (Table \ref{Table:1}; Engagement (NASA TLX)). With Baseline, participants often struggled to concretize shapes using text alone, resulting in designs misaligned with their vision. In contrast, \textit{GenPara} allowed users to clearly express intentions, with the system effectively capturing and generating satisfactory designs (Table \ref{Table:1}; Results Worth Effort (CSI); Figure \ref{fig:13}; Q5, $Mean_{diff} = 1.38$, \(p < 0.010\); Q6, $Mean_{diff} = 1.44$, \(p < 0.050\); Q7, $Mean_{diff} = 1.00$, \(p < 0.001\)). While mental demand showed no significant differences, physical demand was slightly higher for \textit{GenPara} due to its detailed, part-specific editing features (Table \ref{Table:1}; NASA TLX).

\begin{figure*}[h]
  \includegraphics[width=\textwidth]{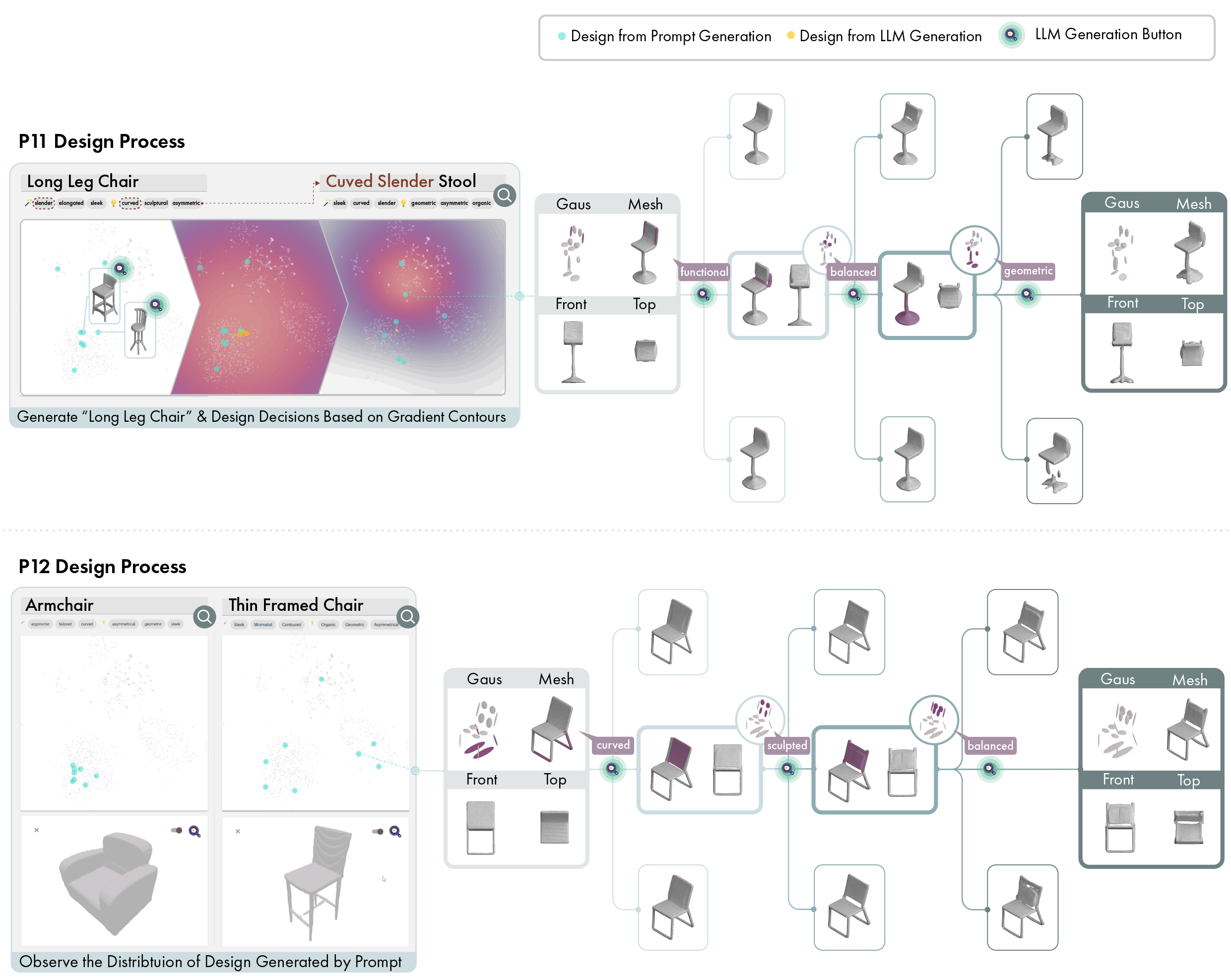}
  \caption{Examples of engrossing into the design generation process of P11 and P12; P11 generates a "long leg chair" and makes design decisions based on gradient contours. P12 explores the distribution of designs generated by prompts and refines the design details progressively. The top and bottom images represent design alternatives generated by the LLM that were not utilized in the design process. Conversely, the central images illustrate design alternatives explicitly refined by the user to achieve their specific goals.}
  \Description{Design generation processes for two participants, P11 and P12, in the GenPara study. P11’s process starts with generating a 'Long Leg Chair' from a prompt and refines it through successive steps, whereas P12 begins with an 'Armchair' and 'Thin Framed Chair', advancing through various stages to refine the designs.}
  \label{fig:14}
\end{figure*}

\begin{figure*}[h]
  \includegraphics[width=\textwidth]{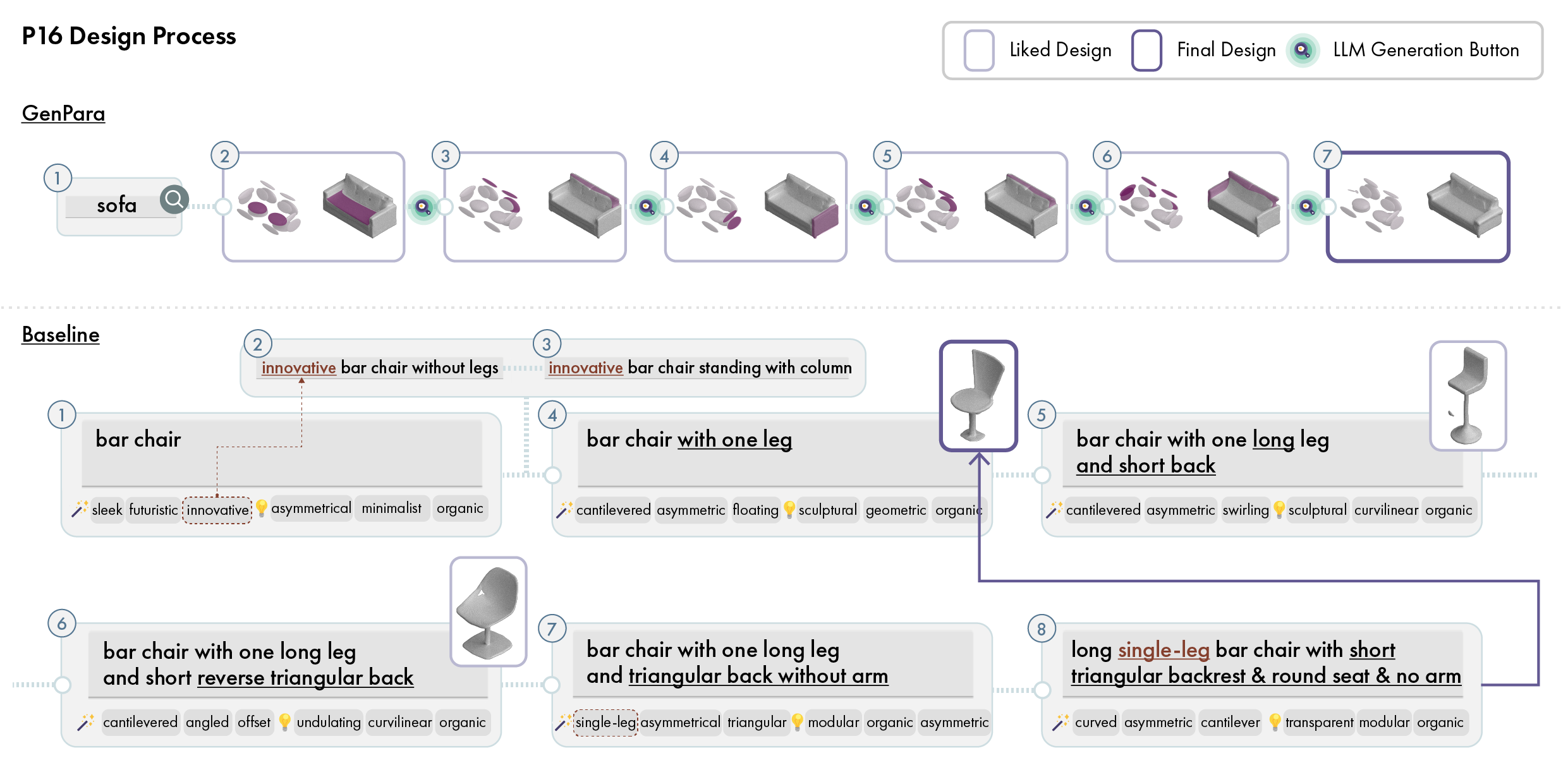}
  \caption{Design concretization process of P16 using \textit{GenPara} and Baseline. The top row shows the iterative design refinement using \textit{GenPara}, starting with a "sofa" and progressing through multiple design modifications. The bottom row illustrates the Baseline process, where the user starts with a "bar chair" and iteratively refines it through text prompts and design adjustments.}
  \Description{Comparative design process for participant P16 using both GenPara and baseline methods. Top sequence shows GenPara's approach starting with a 'sofa' prompt and evolving through several modifications to a refined design. Bottom sequence shows the baseline method starting with a 'bar chair' and progressing through various adjective-enhanced prompts to detailed chair models.}
  \label{fig:15}
\end{figure*}

\begin{figure*}[h]
  \includegraphics[width=\textwidth]{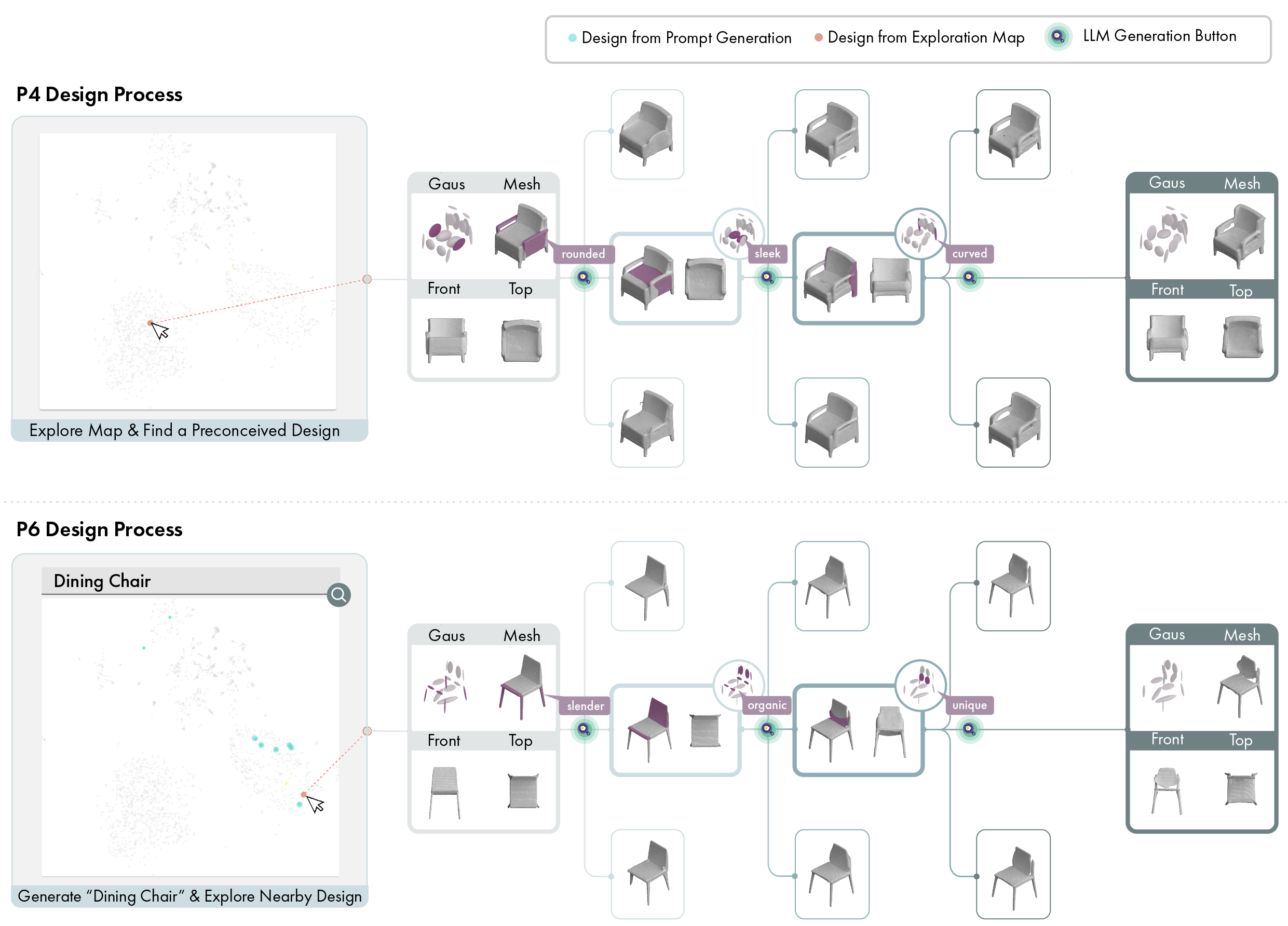}
  \caption{Examples of engrossing into the design generation process of P4 and P6. P4 finds a preconceived design on the Exploration Map and refines it through iterations. P6 generates a "dining chair" and explores nearby designs, progressively refining the details of the chair. The top and bottom images show LLM-generated alternatives not used in the process, while the central images highlight user-refined designs.}
  \Description{Design generation processes for two participants, P4 and P6, in the GenPara study. P4’s process shows a participant exploring a preconceived design and refining it across several stages. P6’s proces details the generation and refinement of a 'Dining Chair' from an initial prompt through several design iterations.}
  \label{fig:17}
\end{figure*}

\textbf{Support for Refining Detailed Goals.} \textit{GenPara} contributed significantly to setting detailed objectives and specifying certain 3D design parts. Unlike Baseline, which encouraged selection from multiple undefined designs, GenPara allowed participants to iteratively refine a single design through part-based editing (P1–P3, P5–P14, P16). Specifically, P14 stated that with Baseline, it was more about selecting the best design among the generated results rather than concretizing a single design. In contrast, P11 mentioned that by directly selecting parts of the design and utilizing designs generated by LLMs in \textit{GenPara}, she realized that she was transforming the initial concept (long leg chair) into a shape with a small arm attached to the back (Figure \ref{fig:14}). P3 mentioned that by starting with a simple search for an "office chair," she was able to specify that the arm should be slanted forward through the LLM. Furthermore, P6 was not satisfied with the result generated by the prompt and started concretizing the design by exploring the surrounding points. He was able to ascertain the direction and achieve specificity in the design when the LLM automatically transformed a back part that was vaguely unsatisfactory (Figure \ref{fig:17}).

\textbf{Support for Generating Shapes Hard to Describe in Text.} Users reported that \textit{GenPara} can generate shapes that are challenging to express clearly in text. They noted that rather than helping to concretize the design shape, they were only recommended various new words through \includegraphics[width=1em]{Figure/alignedAdj.png} aligned and \includegraphics[width=1em]{Figure/diverseAdj.png} diversified prompt suggestions (P2, P3, P5, P12). P10 mentioned, \textit{``When using Baseline, I had to modify prompts multiple times to obtain the chair with extended leg rest), and this process was quite cumbersome and time-consuming.''} Most participants reported that the design generated by prompts did not match their intent or the imagined design, and the designs were not consistent (P1, P3, P4, P7--P12, P14--P16). When P16 was designing a bar chair, the initial generation was satisfactory. However, when attempting to refine it into a single-leg stool with a short back and no arms, the designs did not align with the text, causing P16 to revert to the initial design (Figure \ref{fig:15}). In contrast, when using \textit{GenPara}, P7 said, \textit{``The design alternatives provided by \textit{GenPara} visually represented complex parts effectively, greatly aiding in understanding details that were difficult to explain in text.''} This led to the finalization of a low two-seater with emphasized straight lines, along with a sofa featuring rounded arms and a detailed back. In conclusion, compared with Baseline, \textit{GenPara} allowed for expressing detailed design intentions (Table \ref{Table:1}; Expressiveness (NASA TLX)) with fewer prompt generations (Figure \ref{fig:16}; Prompt Generation Count: $Mean_{Baseline} = 8, Mean_{GenPara} = 2$, \(p < 0.001\)).

\textbf{Detailed Control on 3D Design Parts.} Most participants (P3--P5, P7--P16) mentioned that \textit{GenPara} was helpful in determining the design direction more quickly and accurately. This efficacy arises because, unlike the Baseline system, where participants could directly select chair parts using Gaussian blobs they wanted to edit using only textual descriptions, such as arm, back, seat, and leg. This feature offers greater freedom to specify their designs more concretely. Moreover, it was beneficial for combinations, such as arm and back or back and seat, which allowed for various shape transition suggestions between parts. Participants highlighted that they could specify parts for better shape transitions, such as the front two legs among four or the upper part of the backrest, using Gaussian blobs, thereby conveying their intentions more precisely to the system. For example, P8 wanted to edit only the right armrest to ensure continuity with another sofa in the design brief image, while P13 initially adjusted the front part of the seat and later refined the shape transition by modifying the arm and seat together.

\subsection{RQ3. What is the perceived usefulness of \textit{GenPara} in terms of creativity and efficiency?}

\begin{figure*}[h]
  \includegraphics[width=\textwidth]{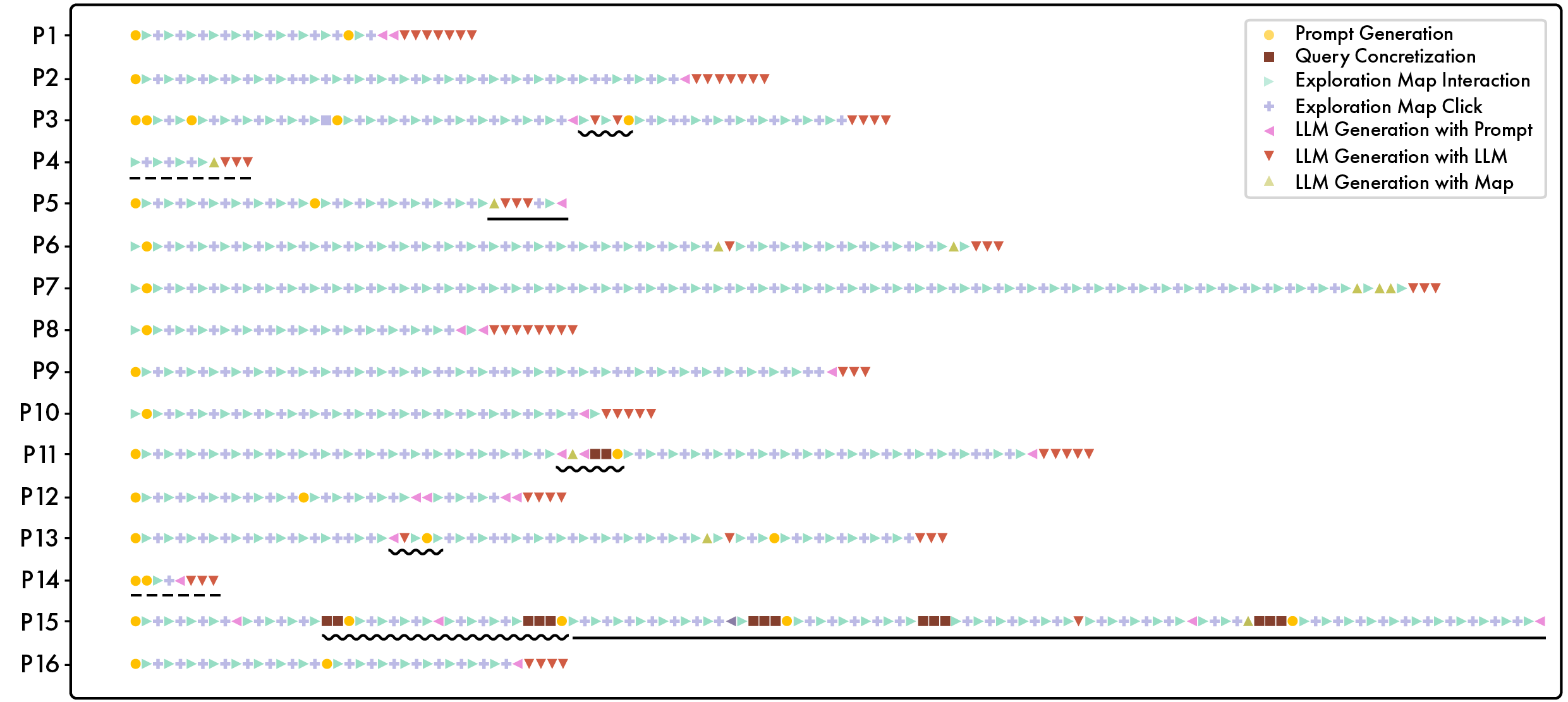}
  \caption{Action log visualization of all participants in \textit{GenPara}, including exploration, prompt generation, and LLM generation. The actual images generated from these design processes are illustrated in Figures \ref{fig:14}, \ref{fig:15}, and \ref{fig:17}. Most participants began their design exploration with vague goals and completed the process by concretizing their objectives through LLM Generation. In the visualization, the dotted underlines represent cases where participants started with a specific design goal based on the design brief, and the wavy underlines indicate cases, where additional prompts were generated to understand the relationship between text and shape parameters using the current design, and the solid underlines, represent cases where participants shifted their goal, leading to the concretization of a new design.}
  \Description{Visualization of all participants’ action logs in GenPara, indicating the variety and frequency of interactions such as prompt generation, query conceptualization, and various forms of map and model generation, depicted through different colored icons and connecting lines.}
  \label{fig:18}
\end{figure*}

We collected responses regarding how each feature of \textit{GenPara} was helpful throughout the design process. Additionally, we analyzed the interaction logs of all participants to further investigate the design process when using \textit{GenPara}. Participants typically began the initial exploration phase with vague design ideas and progressively developed them into concrete designs by utilizing various features of \textit{GenPara} (see Figure \ref{fig:18}). Some participants started exploring specific chair designs and goals (indicated by dotted underlines). Others shifted their design goals during the process (indicated by wavy underlines). Based on these design processes of the participants, we analyzed how each feature of \textit{GenPara} enhances creativity and efficiency in the design process. We briefly summarize the contribution of the key features of \textit{GenPara} to the design process.

\textbf{Input Box Support Aligned and Diversified Prompting.} Most participants used prompts to generate outcomes, particularly during the initial exploration, and simultaneously verified the relationship between the shape parameters and the text they understood. They used prompts to observe how text inputs influenced the overall design shapes. Some participants examined the embedded locations of the outcomes generated through prompts to determine the areas they needed to explore and to understand the relationship between the shape parameters. For example, P8 mentioned using prompts to review the results to verify the detailed relationship between parameters and adjectives obtained. This feature effectively bridged abstract text inputs with the design space, enabling participants to iterate and refine their ideas. Based on the shape parameters they understood, they used prompt generation and the Exploration Map to explore further (solid underlines in Figure \ref{fig:18}).

\textbf{Exploration Map based on UMAP and Bayesian Inference.} The Exploration Map proved instrumental in visualizing diverse design spaces and enabling intuitive navigation of shape parameters. Unlike Baseline, the Exploration Map provided a interactive interface that allowed users to visualize the distribution of design alternatives within the parameter space. For instance, P4 efficiently located a pre-envisioned design without generating additional prompts (see Figure \ref{fig:17}), while P14 generated a design based on his imagination,
confirmed it on the map, and then proceeded to concretization. Some participants shifted their design goals by exploring gradients and nearby points (Figure \ref{fig:18}; solid line) and explored further by generating prompts based on the design parameters they understood (see Figure \ref{fig:18}; wavy line). By offering a clear visual representation of the design space, the Exploration Map enabled participants to grasp the relationship between text inputs and shape parameters. This not only enhanced their understanding but also fostered a more efficient and creative design workflow by facilitating goal refinement and iterative exploration.

\textbf{\includegraphics[width=1em]{Figure/generationButton.png}Generate Design Alternatives with LLMs } Directly selecting parts and generating design alternatives helped participants understand subtle yet distinct shape parameters that were difficult to describe in text. All participants refined their designs within the ROI using \includegraphics[width=1em]{Figure/generationButton.png}, ultimately achieving satisfactory design concretization (Table \ref{Table:1}; Performance (NASA TLX)). Particularly, P3 emphasized that as the design process progressed, the system was good at detecting minor but diverse modifications that were impossible to express in the text. P11 noted that even when they knew which part was unsatisfactory but lacked specific ideas on how to edit it, the system provided various design alternatives, leading to rapid concretization. Participants collectively rated designs generated through part-level modifications more highly than those created solely via prompts, indicating improved efficiency and satisfaction during the design process (Figure \ref{fig:13}; Q9, $Mean_{diff}$ = 1.38, \( p < 0.010 \); Q10, $Mean_{diff}$ = 1.38, \( p < 0.010 \); Q11, $Mean_{diff}$ = 1.00, \( p < 0.050 \)).

\textbf{Design Versioning Tree}. Participants found the hierarchical visualization particularly useful for tracking design changes and understanding how parameter adjustments influenced subsequent outcomes. Most participants noted that the map allowed them to track design progression step-by-step, extend previous designs, and set precise goals for individual parts. P16 stated, “\textit{Visualizing the changes in parameters throughout my design process allowed me to retrospectively review the overall process and assist in decision-making for future steps. It was possible to make modifications to parts while considering shape transitions, allowing for a diverse view.}” Participants highlighted that clear visualization of design stages enabled flexible adjustments and provided a structured overview of their design process.  This capability also facilitated rapid concretization of designs by simplifying the addition or removal of specific details. These insights suggest that the Design Versioning Tree could be a valuable tool for organizing and structuring the design process. Its knowledge-based visualization supported both concrete exploration and detailed refinements, suggesting potential benefits over baseline systems in facilitating iterative development of complex 3D designs.

\section{SUMMARY AND IMPLICATIONS}
Based on the technical evaluation and user study described previously, we aimed to address the implications of \textit{GenPara} as follows. \textit{GenPara} has the potential to enrich our understanding of the design process and shape parameters. By providing designers with a method that facilitates improved efficiency and creativity, \textit{GenPara} introduces novel interactions with 3D GenAI in the design field. These implications can be discussed in three ways.

\subsection{Design Exploration and Concretization with Shape Parameters}
\textit{GenPara} supports the exploration and specification of the design process. Furthermore, \textit{GenPara} enables designers to better understand and articulate their ideas through intuitive visualization and specification of the shape parameters. Exploration based on shape parameters allows designers to compare and assess various design options quickly, leading to more informed design decisions. Additionally, it facilitates the concretization and visualization of design ideas that are difficult for designers to express in text, which is useful in the specification phase of the design process. This support not only facilitates diverse explorations within visually structured design spaces but also has the potential to help designers organize and communicate their ideas more effectively during collaborative processes.

\subsection{Novel Interaction System with LLM in 3D space}
\textit{GenPara} presented a new interaction system in a 3D design space using LLMs. This system offers an intuitive and effective method for designers to create and edit 3D models using text-based descriptions. A core feature of \textit{GenPara} is its ability to explore detailed design modifications based on shape parameters, even without a deep technical knowledge of 3D modeling. This enables designers to clearly and accurately provide instructions on the user-ROIs without an in-depth understanding of complex 3D models. Designers can simplify and accelerate the design process by managing their shape parameters, whereas the interaction between designers and LLMs can support the exploration of diverse design alternatives. Thus, \textit{GenPara} suggests not only the control of design complexity but also the facilitation of creative dialogue between designers and AI.

\subsection{Expanding the Impact of \textit{GenPara} on the 3D Design Process}
\textit{GenPara} is designed to support early-stage designers with a general understanding of the 3D design process but requiring help with exploration, specification, and concretization. However, the user study revealed that \textit{GenPara} is a tool capable of significantly enhancing the 3D design process across various fields and expertise levels. In-depth interviews revealed multiple practical applications where \textit{GenPara} can strengthen the 3D design process with GenAI. Some participants (P3, P8, P10–P12, P14, P15) claimed that \textit{GenPara} enabled them to integrate and visualize ideas alongside traditional sketches and models efficiently. They noted that the LLM-generated design could replace time-consuming sketching or modeling steps. P8 specifically suggested its utility in finalizing designs and refining shape parameters to reduce manual modeling efforts. P5 suggested, “Besides the other three alternatives, it would be nice to directly input text including color, material, etc. for the selected part.” Additionally, the Design Versioning Tree  could support communication in facilitating communication with clients and among designers, helping them express and share design parameters effectively (P1, P2, P4–P7, P16). P16 observed that design is inherently complex and challenging to articulate. He said that \textit{GenPara} could help bridge differences in goals, particularly when objectives are ambiguous. P11 appreciated the automatic visualization and organization of ideas, noting that it aids in structuring design ideation. Furthermore, participants expressed interest in applying \textit{GenPara} to other domains requiring intricate design elements and fine adjustments, such as smartphones, automobiles, architecture, and lighting systems. They emphasized the importance of collaboration with GenAI in the design process, highlighting various possibilities for utilizing \textit{GenPara} across diverse design contexts.

\section{Conclusions}
We introduced \textit{GenPara}, an interactive 3D design editing system designed to enhance the 3D design process by integrating text-conditional shape parameters into a part-aware 3D design. By leveraging LLM, \textit{GenPara} can infer and navigate the ROI through Exploration Map, effectively bridging the gap between GenAI advancements in part assembly and the nuanced requirements of design exploration. The key innovation lies in its ability to generate shape parameters based on textual descriptions of shape parameters. This not only allows for a more tailored design exploration process but also facilitates deeper interaction between designers and GenAI technologies. Moreover, \textit{GenPara} filters and presents design outcomes that align with the envisioned space of the designer and further underscores the capability of the system to provide targeted and relevant design alternatives. While GenPara was designed to support various 3D model categories, this study focused on chair models due to the prevalence of relevant datasets. Expanding to additional domains will require domain-specific fine-tuning to address structural differences and varied interpretations of adjectives, emphasizing the need for enhanced LLM capabilities to effectively accommodate multicategory designs. Although Gaussian blob representations in \textit{GenPara} efficiently model part relationships, future works could integrate comparisons with semantically labeled or segmented parts to further refine workflows and better align transformations with user objectives. Current limitations include challenges with highly irregular shapes or tasks demanding precise localized manual adjustments. Overall, our research elevates designer-GenAI interaction by introducing GenPara, a system that enhances the 3D design editing process through text-conditional shape parameters and their associated design spaces, paving the way for more intuitive and adaptable design systems in the future.

\begin{acks}
This work was supported by National Research Foundation of Korea (NRF) grant funded by the Korean government (MSIT; RS-2023-00208542).
\end{acks}

\bibliographystyle{ACM-Reference-Format}
\bibliography{sample-base}

\end{document}